\documentclass{jpp}
\usepackage{graphicx}
\usepackage{epstopdf, epsfig}

\usepackage[latin1]{inputenc}
\usepackage[T1]{fontenc}
\usepackage{amsmath}
\usepackage{xcolor}

\def\Ha{\mathcal{H}}
\def \Se{\mathcal{S}}
\def\F{\mathcal{F}}
\def\T{\mathcal{T}}
\def\R{\mathbb{R}}
\def\ddd{\textnormal{d}}
\def\ddo{\textnormal{d}^3}
\newcommand{\ddt}[2]{\frac{\textnormal{d}#1}{\textnormal{d}#2}}
\newcommand{\ddp}[2]{\frac{\partial#1}{\partial#2}}
\newcommand{\ddf}[2]{\frac{\delta#1}{\delta#2}}
\newcommand{\vv}[1]{\mathbf{#1}}

\def\p{\partial}
\def\bq{\begin{equation}}
\def\eq{\end{equation}}
\def\bqy{\begin{eqnarray}}
\def\eqy{\end{eqnarray}}
 
\shorttitle{Metriplectic Framework and Dissipative Extended MHD}
\shortauthor{B. Coquinot, P. J. Morrison}

\title{A General Metriplectic Framework with Application to Dissipative Extended Magnetohydrodynamics}

\author{Baptiste Coquinot\aff{1}
  \corresp{\email{baptiste.coquinot@ens.fr}}
 \and Philip J. Morrison\aff{2}\corresp{\email{morrison@physics.utexas.edu}}}

\affiliation{\aff{1}épartement de Physique, École Normale Supérieure, 24 rue Lhomond,  75005, Paris
\aff{2}Department of Physics and Institute for Fusion Studies, University of Texas at Austin, 2525 Speedway, Austin, TX 78712}

\begin{document}

\maketitle

\begin{abstract}
General equations for  conservative yet dissipative (entropy producing) extended  magnetohydrodynamics are derived from  two-fluid theory. Keeping all terms generates unusual cross-effects, such as thermophoresis and a current viscosity that  mixes with the usual velocity viscosity. While the Poisson  bracket of the ideal version of this  model have already been discovered, we determine its  metriplectic counterpart that describes the dissipation. This is done using  a new and general  thermodynamic point of view for deriving dissipative brackets, a means of derivation that is natural for understanding and creating   dissipative dynamics without appealing  to underlying kinetic theory orderings. Finally the  formalism is used to study dissipation in the Lagrangian variable picture where,  in the context of extended  magnetohydrodynamics,  nonlocal dissipative brackets naturally emerge.
\end{abstract}

%\keyword{Fluid Dynamics; Hamiltonian Dynamics; Non-Equilibrium Thermodynamics; Plasma Physics; Dissipative Extended Magnetohydrodynamics; Metriplectic Brackets}

%Main text
\section{Introduction}

\subsection{Background}

It is well known that the Hamiltonian dynamics of discrete and continuum systems may be written in terms of  Poisson brackets (\cite{Landau60book, Morrison98}) with  Hamiltonians. Such systems describe the evolution of a point in a phase space that may be finite-dimensional, the case for discrete systems,  or infinite-dimensional, the case for continuum systems. The Poisson bracket  $[f,g]$ is a bilinear operation on the set of smooth phase space functions  $f,g$ or  functionals  ($0$-forms) that maps  phase space to the real numbers. This set includes all the physical observables of interest.  The Poisson bracket is also  skew-symmetric, is a derivation,  and satisfies the  Jacobi identity; it generates the dynamics for any phase space function according to
\[
\frac{df}{dt}= [f,\mathcal{H}]\,,
\]
where  the Hamiltonian $\mathcal{H}$, which is usually the energy, plays a special role.  In the canonical case the Poisson bracket is nondegenerate and the Jacobi identity is equivalent to  the  associated symplectic 2-form being closed, while in the noncanonical case degeneracy gives rise to Casimir invariants,   particular functionals having vanishing Poisson brackets with all functionals, {\it i.e.}, Casimirs $C$ satisfy $[f,C]=0\  \forall f$.  The rich geometry of a phase space with a defined Poisson bracket, which includes symplectic and Poisson geometry, is  intellectually very interesting and allows  for  better insight.  Moreover,  it is of  practical value for understanding spectra, perturbation theory,  and the construction of numerical algorithms (see,  \textit{e.g.}, \cite{Salmon83,hagstrom,gempic,vanneste}).

Less known is the fact  that dissipative dynamics can also emerge from brackets \citep{pjmK82,Morrison84,Kaufman84,Morrison84b,Grmela84,Morrison86}  and the entropy $\Se$ rather than  the Hamiltonian $\Ha$ may serve as the generating function.   (See \cite{Grmela97i,Edwards98,Morrison09,pjmK14,Materassi18, Gay-Balmaz17i, Gay-Balmaz17ii, Eldred18,Grmela97ii} for a selection of more recent theoretical work,  and \textit{e.g.}\  \cite{eero,pjm17,Bressan18} for recent numerical algorithms based on  bracket dissipative structure.)   Given that physical models generally contain both Hamiltonian  and dissipative parts, we would like to use both kinds of brackets to get the complete dynamics. To this end we introduce a free energy $\F=\Ha-\T\Se$, where $\T$ is a Lagrange multiplier, interpreted as a generalized temperature (which is a uniform constant, in opposition to the physical temperature of the system $T$). This generalization is natural both because  the free energy has a physical interpretation and  because the entropy is   a Casimir invariant of the Poisson bracket. The dynamics then becomes for any functional $f$ of the system, 
\[
\ddt{f}{t}=\big[[f,\F]\big]\,, 
\]
where $\big[[f,g]\big]$ is an inclusive  bracket, defined as the difference between a Poisson bracket, denoted $\{f,g\}$, and a dissipative bracket, denoted $(f,g)$. Since the energy is preserved and the entropy increases with time, $\mathcal{F}$ is a thermodynamic potential. Then, an equilibrium is given by $\delta\mathcal{F}=0$,  where $\delta$ means the functional variation, which will be formally defined later. 
 
 An  interesting property can already  be proven. Upon  denoting  by $\sigma$ the entropy density,  assuming  that this variable appears in the Hamiltonian only through an internal energy density $u_{\textnormal{\tiny{Vol}}}$,  and making use of the usual  thermodynamic definition of  the local temperature, $T= \p{u_{\textnormal{\tiny{Vol}}}}/\p{\sigma}$, we find the variation of $\mathcal{F}$ induced by  a perturbation of the entropy $\delta\sigma$ gives $\delta{\F}/\delta{\sigma}=T-\T$.  Therefore, at equilibrium, the temperature is uniform and equals the Lagrange multiplier $\mathcal{T}$, which validates our interpretation of this constant.

To be compatible with thermodynamics, there are requirements. The second law of thermodynamics is assured if  the dissipative bracket has a nonnegative symmetrical bilinear form (we assume $\T$ to be nonnegative, which is consistent with its equilibrium interpretation). Then, it almost defines a metric. The first law requires the conservation of the Hamiltonian, \textit{i.e.}\  $(\Ha,\F)=0$. A stronger assumption is to require the degeneracy,  \textit{i.e.}\   $(\Ha,f)=0$ for any functional $f$. The situation is then symmetrical with the fact that $\Se$ is a Casimir of the Poisson bracket. If such an  assumption is fulfilled,  the dissipative bracket is called metriplectic  and metriplectic dynamics of any functional $f$ is given by 
\begin{equation*}
\ddt{f}{t}=\big[[f,\F]\big]=\{f,\Ha\}+\T(f,\Se)\,.
\nonumber
\end{equation*}
The two principles of thermodynamics are then fulfilled by construction: 
\begin{eqnarray}
\ddt{\Ha}{t}&=&\{\Ha,\Ha\}+\T(\Ha,\Se)=0\,; 
\nonumber\\
 \ddt{\Se}{t}&=&\{\Se,\Ha\}+\T(\Se,\Se)=\T(\Se,\Se)\geq0\,.
 \nonumber
\end{eqnarray}

The formalism above with the properties of symmetry and degeneracy  first appeared in \cite{Morrison84,Morrison84b}, with the terminology metriplectic introduced in \cite{Morrison86}.  Several examples were given in these early works.  Later it was  called generic in \cite{Grmela97i}.

For a given physical system, it remains to determine the brackets.  For fluid-like theories, the Poisson brackets naturally come from canonical brackets in terms of Lagrangian position and  momentum variables, which can then be transformed into the usual Eulerian variables (see \textit{e.g.}\  \cite{Morrison98}).   Poisson brackets for many models of plasma physics exist in the literature, including those for   magnetohydrodynamics (\cite{pjmG80,Morrison09conf}), relativistic magnetohydrodynamics \cite{DAvignon15} and extended magnetohydrodynamics (\cite{DAvignon16}). They are also known in other  subjects of physics, \textit{e.g.},  in geophysical fluids (\cite{Bannon03}) and elasticity (\cite{Edwards91}). Fewer metriplectic brackets are known; however, they have been discovered for fluids with viscosity and thermal diffusion (\cite{Morrison84b}),   elasticity (\cite{Edwards91}),  and magnetohydrodynamics (\cite{Materassi12}). A comprehensive Lagrangian based approach, as opposed to our  bracket approach,  is given for $n$-fluid models with chemical reactions and  general multicomponent fluids with irreversible processes in \citet{Eldred18,Eldred20}.

\subsection{Extended Magnetohydrodynamics: Model}
\label{ssec:XMHDmodel}

In this paper, we will mostly focus on extended magnetohydrodynamics. Extended magnetohydrodynamics may be derived from two-fluid theory, where ions and electrons are treated as distinct fluids. From this model, it is possible to get new equations in the usual variables,   the center-of-mass velocity $\vv{v}$ and the electrical current $\vv{j}$ (\cite{Lust59, Kampen67book}) (see also \cite{Lingam16, Charidakos14}). To simplify the equations,  two assumptions are made:  quasineutrality, \textit{viz.}\ that the densities of electrons and ions are assumed equal,  and that  the ratio of masses of electrons and ions, $\mu={m_e}/{m_i}$,  is small so that one can expand in powers of $\mu$. With these  assumptions, one gets a generalized equation of motion  for the velocity and a generalized Ohm's law.  The model restricted to zeroth order  in $\mu$ is called Hall magnetohydrodynamics (HMHD),  while retaining first order  terms  produces what has been called extended magnetohydrodynamics (XMHD). 

The equations of XMHD  are expressed in terms of the  variables $(\rho, \sigma, \sigma_e, \vv{v}, \vv{j}, \vv{E}, \vv{B})$,  which are respectively the mass density, the total entropy density, the electron entropy density, the center-of-mass velocity, the electrical current and the electric and magnetic fields. If our plasma is confined in a domain $\Omega$, the three Eulerian scalars are functions from $\Omega\times \R\longrightarrow \R$, while the four vector fields are functions from $\Omega\times \R\longrightarrow \textnormal{T}\Omega$, where $\textnormal{T}\Omega$ stands for the tangent bundle of the manifold $\Omega$  (here taken to be simple, {\it e.g.}, a three torus). For simplicity, we define 
$\chi={m_i}/{e}$, with $e$ being the charge of both electrons and ions, and choose units such that $\mu_0=\varepsilon_0= c=1$.
The three scalar fields  satisfy the conservation laws, 
$$
\partial_t\rho+\nabla\cdot(\rho\vv{v})=0,  \hspace{1.2cm} \partial_t\sigma+\nabla\cdot(\sigma\vv{v})=0,  \hspace{1.2cm} \partial_t\sigma_e+\nabla\cdot(\sigma_e\vv{v}_e)=0\,,
$$
where at order one in $\mu$, the electron velocity $\vv{v}_e=\vv{v}-(1-\mu)\chi\, {\vv{j}}/{\rho}$. The center of mass velocity satisfies the  momentum conservation law,  
\[
\ddt{\vv{v}}{t}= \p_t \vv{v}  + \vv{v}\cdot \nabla \vv{v}= -\frac{1}{\rho}\nabla P +\frac{\vv{j}}{\rho}\times\vv{B}-\mu\chi^2\, \frac{\vv{j}}{\rho}\cdot\nabla\left(\frac{\vv{j}}{\rho}\right)\,,
\]
where  the use of  $\mathrm{d}/\mathrm{d}t$ for the advective derivative should be clear from context, while its counterpart is the generalized Ohm's law, 
\bqy
\vv{E}+\vv{v}\times \vv{B}&=&\chi\frac{\vv{j}}{\rho}\times\vv{B}-\frac{\chi}{\rho}\nabla P_e
\nonumber\\
&+&\mu\frac{\chi^2}{\rho}\left[\ddp{\vv{j}}{t}+\nabla\cdot(\vv{v}\otimes \vv{j}+\vv{j}\otimes \vv{v})\right]-\mu\chi^3\frac{\vv{j}}{\rho}\cdot\nabla\left(\frac{\vv{j}}{\rho}\right)\,.
\nonumber
\eqy
Finally, the electromagnetic variables are linked by the pre-Maxwell equations, 
\bqy
\partial_t\vv{B}+\nabla\times\vv{E}=0,  \hspace{1.2cm} \nabla\times\vv{B}=\vv{j},  \hspace{1.2cm} \nabla\cdot\vv{B}=0\,.
\nonumber
\eqy
To close the system, one must specify the internal energies per unit mass for both ions, $u_i(\rho_i, \sigma_i)$,  and  electrons,  $u_e(\rho_e, \sigma_e)$, where $\rho_i \approx (1-\mu)\rho$, $\rho_e \approx \mu\rho$ and  $\sigma_i=\sigma-\sigma_e$. Then, the pressures are determined by 
$$
P_i=\rho_i^2\ddp{u_i}{\rho_i}\,,  \qquad  P_e=\rho_e^2\ddp{u_e}{\rho_e}\,, \qquad 
\mathrm{and}\qquad  P= P_i+P_e\,.
$$

One can also simplify the XMHD equations by eliminating the variables $\vv{j}$ and $\vv{E}$, which is  indeed useful since the Ohm's law and Ampere's  equation  are not evolution equations but constraint equations. Then, the phase space should be a submanifold of the $(\rho, \sigma, \sigma_e, \vv{v}, \vv{j}, \vv{E}, \vv{B})$ vector space. Thus, we are able to reduce these variables to get an easier phase space. To this end, it is useful to define  a new variable, $\vv{B^*}$, that we will refer to here as the  drifted magnetic field, 
\bq
\vv{B^*}:= \vv{B}+\mu\chi^2\, \nabla\times\left(\frac{\vv{j}}{\rho}\right)
=\vv{B}+\mu\chi^2\, \nabla\times\left(\frac{1}{\rho}\nabla\times\vv{B}\right)\,.
\label{Bstar}
\eq
This variable first appeared in \citet{pjmLT15}.
Physically, this drift comes from the difference of inertia between ions and electrons. While the velocity represents  mostly the movement of ions, the \textit{frozen-in} property of the magnetic field  (\cite{alfven}) (also see \textit{e.g.}\ \cite{Kampen67book}) is related to the dynamics of electrons. This creates a drift between the velocity and the magnetic flux, which is taken into account in this drifted magnetic field. 

The Ohm's law and the pre-Maxwell's equations then reduce to (\cite{DAvignon16}) 

\bqy
\partial_t\vv{B^*}&=&\nabla\times\left[\vv{v}\times\vv{B^*}-\frac{1}{\rho}\left(\nabla\times\vv{B}\right)\times\vv{B^*}+\mu\frac{\chi}{\rho}\left(\nabla\times\vv{B}\right)\times\left(\nabla\times\vv{v}\right)\right]
\nonumber\\
&&\hspace{5.8cm}
-\frac{\chi}{\rho^2}\left(\nabla P_e \times \nabla \rho\right)\,.
\nonumber
\eqy

\subsection{Extended Magnetohydrodynamics: Geometry}
\label{ssec:XMHDham}

After the reduction of removing $\vv{E}$ and $\vv{j}$, the phase variables can be chosen to be $(\rho, \sigma, \sigma_e ,\vv{m}, \vv{B^*})$,  where $\vv{m}= \rho\vv{v}$ is the momentum density and $\vv{B^*}$ is constrained to be a divergence-free vector field.  However, the divergence-free constraint will be fulfilled if it  is  initially true, because it turns out the dynamics will  propagate it. Then, we define the local phase space at a point $\vv{x}\in\Omega$ as 
\[
\Phi_{\vv{x}}= \R_\rho\times \R_\sigma\times\R_{\sigma_e}\times \left(\textnormal{T}_{\vv{x}}\Omega\right)_\vv{m}\times \left(\textnormal{T}_{\vv{x}}\Omega\right)_\vv{B^*}
\]
and the global phase space $\Phi$ as the sections of the bundle 
\[
\amalg_{\vv{x}\in\Omega}\Phi_{\vv{x}}\longrightarrow \Omega\,.
\]
That is, a point of the global phase space gives  an element of the local phase space for each spatial position, which describes uniquely the state of our system. We then define a functional as a  map  from $\Phi\longrightarrow \R$ (or $\R^3$ for vectors), a bracket as a bilinear operator $\R^\Phi\times \R^\Phi\longrightarrow \R^\Phi$ (where $\R^\Phi$ denotes the map  from $\Phi$ to $\R$), that fulfills the Leibniz rule, the variation of a functional $f$ as $\delta f:\Phi\times\Phi\longrightarrow \R$ given by 
\[
\delta f(\varphi, \delta\varphi)=\textnormal{lim}_{\epsilon\rightarrow 0}\frac{f(\varphi+\epsilon\delta\varphi)-f(\varphi)}{\epsilon}\,,
\]
which can be viewed  as directional derivative of $f$ at $\varphi$ in the direction $\delta\varphi$,
and the functional derivative  of $f$ at a point $\delta\varphi\in\Phi$, denoted $\delta{f}/{\delta \varphi}$, as the functional, when it exists, that satisfies, 
\[
  \delta f(\varphi, \delta\varphi)=\int_{\Omega}\ddf{f}{\varphi}\delta\varphi\,,
\] 
 for any $\varphi\in\Phi$, where the volume element (\textit{e.g.} $\textnormal{d}^3x$) will not be stated when there is no likelihood of confusion. We consider a particular physical path in the phase space,  parametrized by the time $\R$, and then functionals may be seen as functions of time. Let us also notice that a function on a local phase space $g:\Phi_\vv{x}\longrightarrow \R$,  where $\vv{x}\in\Omega$, may be seen as a functional $g^x: \varphi\in\Phi\longrightarrow \int_\Omega g(\varphi(\vv{y}))\delta_\Omega(\vv{x}-\vv{y})\textnormal{d}^3y$,  where $\delta_\Omega$ is the Dirac distribution on $\Omega$ (and assuming that we can define $g$ over any local phase space, which is natural in practice). Without changing notations we identify $g$ with $g^x$.
  
Finally, one must define the important functionals $\Se$ and $\Ha$. First,  of course, we have the  total entropy, 
\bq
\Se=\int_\Omega \sigma\,.
\label{entropy}
\eq
Second, the energy per unit volume contains the kinetic energy, the  internal energy, the magnetic energy, and also  the kinetic energy of the electrons (\cite{pjmK14}). We denote the global internal energy per unit mass $u=({\rho_i u_i+\rho_e u_e})/{\rho}=(1-\mu)u_i+\mu u_e$. All together, this gives the total energy density, 
%simplified in the means that it has sense only in an integral
\bq
\varepsilon= \frac{1}{2}\rho|\vv{v}|^2+ \rho u+ \frac{1}{2}|\vv{B}|^2+\frac{1}{2}\mu\frac{\chi^2}{\rho}|\vv{j}|^2=\frac{|\vv{m}|^2}{2\rho}+ \rho u+ \frac{1}{2}\vv{B}\cdot\vv{B^*}\,, 
\label{xmhdE}
\eq
where use has been made of (\ref{Bstar}) and in the last equality a total divergence has been dropped.  Thus this and other equalities involving integrands should  be interpreted modulo a surface term. 
Then, the Hamiltonian is 
\bq
\Ha= \int_\Omega\varepsilon=\int_\Omega\left(\frac{|\vv{m}|^2}{2\rho}+\rho u+ \frac{1}{2}\vv{B}\cdot\vv{B^*}\right)\,.
\label{hamiltonian}
\eq

For XMHD, the following Poisson bracket on \textit{e.g.}  functionals $f,g$, which  was  first given  in \cite{hamdi}  based on the earlier work of \cite{pjmK14,pjmLT15}, together with the Hamiltonian of (\ref{hamiltonian}) produces the equations of motion:   
\begin{eqnarray}
&&\hspace{-.5cm}\{f,g\}=\int_\Omega \ddo{y}\, \Bigg[
\rho\frac{\delta f}{\delta \vv{m}(\vv{y})}\cdot\nabla\left(\frac{\delta g}{\delta \rho(\vv{y})}\right)-\rho\frac{\delta g}{\delta \vv{m}(\vv{y})}\cdot\nabla\left(\frac{\delta f}{\delta \rho(\vv{y})}\right)
\nonumber\\
	&+&\sigma\frac{\delta f}{\delta \vv{m}(\vv{y})}\cdot\nabla\left(\frac{\delta g}{\delta \sigma(\vv{y})}\right)-\sigma\frac{\delta g}{\delta \vv{m}(\vv{y})}\cdot\nabla\left(\frac{\delta f}{\delta \sigma(\vv{y})}\right)
	\nonumber\\
	&+&\vv{m}\cdot\left(\frac{\delta f}{\delta \vv{m}(\vv{y})}\cdot\nabla\left(\frac{\delta g}{\delta \vv{m}(\vv{y})}\right)\right)-\vv{m}\cdot\left(\frac{\delta g}{\delta \vv{m}(\vv{y})}\cdot\nabla\left(\frac{\delta f}{\delta \vv{m}(\vv{y})}\right)\right)
	\nonumber\\
	&+&\vv{B^*}\cdot\left(\frac{\delta f}{\delta \vv{m}(\vv{y})}\cdot\nabla\left(\frac{\delta g}{\delta \vv{B^*}(\vv{y})}\right)\right)-\vv{B^*}\cdot\left(\frac{\delta g}{\delta \vv{m}(\vv{y})}\cdot\nabla\left(\frac{\delta f}{\delta \vv{B^*}(\vv{y})}\right)\right)
	\nonumber\\
	&-&\frac{\delta f}{\delta \vv{m}(\vv{y})}\cdot\left(\vv{B^*}\cdot\nabla\left(\frac{\delta g}{\delta \vv{B^*}(\vv{y})}\right)\right)+\frac{\delta g}{\delta \vv{m}(\vv{y})}\cdot\left(\vv{B^*}\cdot\nabla\left(\frac{\delta f}{\delta \vv{B^*}(\vv{y})}\right)\right)
	\nonumber\\
	&-&\frac{c\chi}{\rho}\left((1+\mu)\vv{B^*}-\mu\chi\nabla\times\left(\frac{\vv{m}}{\rho}\right)\right)
	\nonumber\\
	&&\hspace{2cm}\times \left(\nabla\times\left(\frac{\delta f}{\delta\vv{B^*}(\vv{y})}\right)\right)\times\left(\nabla\times\left(\frac{\delta g}{\delta\vv{B^*}(\vv{y})}\right)\right)\Bigg]\,.
\label{hamdi}	
\end{eqnarray}
This bracket is clearly skew-symmetric in $f$, $g$ and it was shown by direct calculation in \cite{hamdi}  to satisfy the  Jacobi identity.  A much simplified proof of the Jacobi identity along with some remarkable connections to other models was obtained in  \cite{pjmLM15} and the bracket of (\ref{hamdi}) was derived from a Lagrangian variable action functional in  \cite{DAvignon16}. 
 
 Lastly, we note that strong boundary conditions are assumed such that all needed integrations by parts produce no boundary terms.   In this paper, we will not consider any boundary effect on the brackets.

\subsection{Development - Overview}

Given the model and Hamiltonian structure of sections \ref{ssec:XMHDmodel}  and \ref{ssec:XMHDham}, respectively, it remains to discuss dissipation,  the main content of our paper.  The dynamical variables,  the phase space,  will remain the same, but the evolution equations will obtain new dissipative terms  generated by a metriplectic bracket.  A  general form with several  dissipative effects  will be obtained by  what amounts to a purely thermodynamic means.  We emphasize that  our approach  differs from the very large plasma literature that yields fluid transport properties by appealing to particular kinetic theory orderings, {\it e.g.}, the classic reference of \citet{braginskii}   (see also \citep{kulsrud}) and  many subsequent highly detailed works.

In section \ref{sec:DXMHD} we start again from two-fluid theory, with general forms of thermal and viscous dissipative terms, to obtain  dissipative XMHD. We will also consider cross-terms and look at their  effect. For example, our model will include thermophoresis, and we will also discover  a new current viscosity. This new viscosity will allow new cross effects between the velocity and magnetic field evolutions that may be of higher derivative orders while remaining linear.  

In section \ref{sec:bracket} we will examine the dissipative brackets from a purely thermodynamic point of view, and find from this new perspective the brackets of hydrodynamics. Next we will introduce another set of variables that appear more natural for constructing the metriplectic bracket and then present a systematic way to derive general  dissipative brackets for fluid-like systems.

In section \ref{sec:detbracket} we will determine the complete brackets of dissipative XMHD. Keeping all cross effects, the dissipative bracket is given by equation  (\ref{dispbkt})  with the various phenomenological coefficients including cross effects described there.

In section \ref{sec:lag}, we discuss the Lagrangian variable picture of this model. While the Hamiltonian dynamics is well-known in terms of these variables,  dissipation is usually considered only in the Eulerian picture.  Consequently, we complete the picture by examining general forms of dissipation  in the Lagrangian variable picture, and describe the transformation to the Eulerian picture as an example of metriplectic reduction.  

Finally, in section \ref{sec:conclusion} we conclude this work.

\section{From Two-Fluid Theory to Dissipative Extended Magnetohydrodynamics}
\label{sec:DXMHD}

\subsection{Two-Fluid Theory}

In the two-fluid model, one considers the ions and the electrons as two distinct fluids with two velocity fields, $\vv{v}_i$ and $\vv{v}_e$, two mass densities, $\rho_i$ and $\rho_e$, and two pressures, $P_i$ and $P_e$. In addition one  has the individual mass conservation laws,  	
\bq
\partial_t\rho_i+\nabla\cdot(\rho_i\vv{v}_i)=0 
\qquad \mathrm{and}\qquad
 \partial_t\rho_e+\nabla\cdot(\rho_e\vv{v}_e)=0\,.
\label{density}
\eq
The quasineutrality assumption states, to first order  in $\mu$,  
\[
\rho_i\approx(1-\mu)\rho \qquad \mathrm{and}\qquad  \rho_e\approx \mu\rho\,,
\]
where $\rho=\rho_i+\rho_e$.  The fluid equations for these variables are then coupled via the  pre-Maxwell equations. 

Conductivity arises from collisions between electrons and ions, and these are  modeled by an exchange term in the momentum equation proportional to the relative velocity (see \textit{e.g.}\ \cite{Kampen67book}). This term  will express the resistivity of  Ohm's law, with the bonus of a physical interpretation at this level. Plus, we know that in the one-fluid theory,   Ohm's law has a tensorial phenomenological constant, like what may occur for  Fourier's law of heat conduction. From a thermodynamic point of view there could also be cross-terms between heat and electrical conduction (\cite{Groot84book}); consequently,  for  generality, we will  add such cross terms in our two-fluid theory.  To conserve the total momentum these appear with opposite signs.  

A decision about  temperature needs to be made.  Although in many plasmas the electrons and ions have not relaxed, we will assume a common temperature.  This is done for simplicity in this paper, mostly to not further complicate the presentation.  This assumption allows us to drop the $\sigma_e$ variable, as will be discussed  later, but  generalizing   the bracket by dropping this  hypothesis is possible.

We assume that both fluids have their individual viscosities,  which together will generate a one-fluid viscosity. This assumption naturally produces additional viscosities:   a current viscosity and cross-effect viscosities, which do not appear to have been heretofore explored.  Physically, all viscosities can be traced back to  collisions between particles of the fluids. We are  also able to add cross viscosity at the outset  between the fluids, but this only alters the phenomenological coefficients that appear in our final theory. The terms we obtain can be interpreted as arising from collisions between ions and electrons of an higher order than the usual exchange terms. 

Given the above assumptions, the two-fluid equations 
\citep[cf.\ {\it e.g.}][]{kulsrud} become
\begin{eqnarray}\label{vi}
	\hspace{-0.5cm}\rho_i\left(\partial_t \vv{v}_i+\vv{v}_i\cdot\nabla\vv{v}_i\right)=&-\nabla P_i+\chi^{-1}\rho_i\left(\vv{E}+\vv{v}_i \times\vv{B}\right)-\frac{\rho^2}{\chi^2}\eta_{jj}(\vv{v}_i-\vv{v}_e) \nonumber\\
	&-\frac{\rho}{\chi}\eta_{jT}\nabla T+\nabla\cdot\left[\Lambda_{ii}\nabla\vv{v}_i+ 
	\Lambda_{ie}\nabla\vv{v}_e\right]
\end{eqnarray}
 and 
\begin{eqnarray}\label{ve}
	\hspace{-0.5cm}\rho_e\left(\partial_t \vv{v}_e+\vv{v}_e\cdot\nabla\vv{v}_e\right)=&-\nabla P_e-\chi^{-1}\rho_i\left(\vv{E}+\vv{v}_e \times\vv{B}\right)+\frac{\rho^2}{\chi^2}\eta_{jj}(\vv{v}_i-\vv{v}_e) \nonumber\\
	&+\frac{\rho}{\chi}\eta_{jT}\nabla T+\nabla\cdot\left[\Lambda_{ei}\nabla\vv{v}_i+\Lambda_{ee}\nabla\vv{v}_e\right]\,,
\end{eqnarray}
where we have introduce six phenomenological coefficients. Two conductivities, electrical with $\eta_{jj}$ and thermic with $\eta_{jT}$,  and four viscosity coefficients, for ions $\Lambda_{ii}$,  electrons $\Lambda_{ee}$,  and cross effects $\Lambda_{ie}$ and $ \Lambda_{ie}$,  where the latter are symmetric because of the Onsager relations.    The conductivities  are in general 2-tensors, \textit{i.e.},  matrices on $\textnormal{T}_\vv{x}\Omega$ at each point $\vv{x}\in\Omega$, whereas the viscosities  are 4-tensors,  \textit{i.e.},  matrices on the vector space of matrices on $\textnormal{T}_\vv{x}\Omega$.    Thus, \textit{e.g.},   the $b$th-component of the term $\nabla\cdot\left[\Lambda_{ii}\nabla\vv{v}_i\right])$ in cartesian tensor notation, where repeated indices are summed,  is $\p_a[(\Lambda_{ii})_{abcd}\p_c ({v}_i)_d]$, with  $a,b,c,d\in\{1,2,3\}$. Other tensor expressions here and henceforth should be interpreted similarly. We choose some constants with these coefficients, which change nothing since one may define the phenomenological coefficients another way, but will simplify the calculations.

 For generality we allow the phenomenological coefficients to be arbitrary: they may be  general tensors  that  depend on the phase space variables and position. 

Lastly,  thermodynamics gives two equations, the thermodynamic identities, 
\[
T\ddd s_i=\ddd u_i-\frac{P_i}{\rho_i^2}\, \ddd\rho_i \qquad \mathrm{and} \qquad
T\ddd s_e=\ddd u_e-\frac{P_e}{\rho_e^2}\, \ddd\rho_e,
\]
where $s_i$ and $s_e$ are the specific entropies of ions and electrons,  while $u_i$ and $u_e$ are the specific internal energies of ions and electrons. Later, we will prefer to use the densities, which are related to the specific entropies and internal energies according to  $\sigma_\alpha=\rho_\alpha s_\alpha$ and $u_{\alpha, \textnormal{\tiny{Vol}}}=\rho_\alpha u_\alpha$ for $\alpha\in\{i, e\}$.

\subsection{Toward One-Fluid Theory}

Next we define the center-of-mass velocity $\vv{v}$, the electrical current $\vv{j}$,  and the total pressure $P$ by 
$$
\vv{v}= (\rho_i\vv{v}_i+\rho_e\vv{v}_e)/\rho  =\vv{v}_i+\frac{\mu}{1+\mu}(\vv{v}_e-\vv{v}_i)\,,  
\quad \vv{j}= \rho_i(\vv{v}_i-\vv{v}_e)/\chi\,,  \quad  P= P_i+P_e\,.
$$
We may also write, 
\[
\vv{v}_i\approx \vv{v}+\mu\chi\,{\vv{j}}/{\rho} \quad \mathrm{and}\quad \vv{v}_e\approx \vv{v}-\chi\, {\vv{j}}/{\rho}\,.
\]

Given the above change of variables and the quasineutrality assumption,  equations (\ref{density}) imply the one-fluid mass conservation law,
\[
\partial_t\rho+\nabla\cdot\left(\rho\vv{v}\right)=0\,.
\]
The one-fluid velocity equation and  Ohm's law  follow from the sum and difference of equations (\ref{vi}) and (\ref{ve}). The more difficult part of  this computation is to manage the nonlinear terms. However, these terms are purely dynamical and not dissipative, and they have  already been derived. For the detailed computation of these nonlinear terms, see \citet{Lust59, DAvignon16}.
Summing equations (\ref{vi}) and (\ref{ve}) gives
\bqy
\rho\left[\partial_t\vv{v}+\vv{v}\cdot\nabla\vv{v} 
+\mu\chi^2\,\frac{\vv{j}}{\rho}\cdot\nabla\left(\frac{\vv{j}}{\rho}\right)\right]
&=&\vv{j}\times\vv{B}-\nabla P
\nonumber\\
&+&
\nabla\cdot\left[\Lambda_{vv}\nabla\vv{v}+\Lambda_{vj}\nabla\left(\frac{\vv{j}}{\rho}\right)\right],
\nonumber
\eqy
where 
\[
\Lambda_{vv}= \Lambda_{ii}+\Lambda_{ie}+\Lambda_{ei}+\Lambda_{ee}
\]
and 
\[
 \Lambda_{vj}=-\chi\left(\Lambda_{ie}+\Lambda_{ee}\right)+\mu\chi\left(\Lambda_{ii}+\Lambda_{ei}\right).
\]
We can also define the viscosity 2-tensor
\[
\Pi_v= \Lambda_{vv}\nabla\vv{v}+\Lambda_{vj}\nabla\left(\frac{\vv{j}}{\rho}\right)\,.
\]
 
The independent combination $\mu(1-\mu)\frac{\chi}{\rho} \times(\ref{vi}) - (1-\mu)\frac{\chi}{\rho}\times(\ref{ve})$ gives the following:
\bqy
\label{ohm}
&&\mu\frac{\chi^2}{\rho}\left[\ddp{\vv{j}}{t}+\nabla\cdot(\vv{v}\otimes \vv{j}+\vv{j}\otimes \vv{v})-\chi\vv{j}\cdot\nabla\left(\frac{\vv{j}}{\rho}\right)\right] + \eta_{jj}\,\vv{j}+\eta_{jT}\nabla T\\
&&\hspace{1.6cm} =\vv{E}+\left(\vv{v}-\chi\frac{\vv{j}}{\rho}\right)\times \vv{B}+\frac{\chi}{\rho}\nabla \left(P_e-\mu P\right)+\frac{1}{\rho}\nabla\cdot\left[\Lambda_{jv}\nabla\vv{v}+\Lambda_{jj}\nabla\left(\frac{\vv{j}}{\rho}\right)\right]\,,
\nonumber
\eqy
where 
\bq
\Lambda_{jv}= -\chi\left(\Lambda_{ei}+\Lambda_{ee}\right)+\mu\chi\left(\Lambda_{ii}+\Lambda_{ie}\right)
\nonumber
\eq
 and 
\bq
\Lambda_{jj}= \chi^2\Lambda_{ee}-\mu\chi^2\left(\Lambda_{ii}-\Lambda_{ie}-\Lambda_{ei}\right)\,.
\nonumber
\eq
To maintain a consistent ordering we let $P_e-\mu P\rightarrow P_e$ by absorbing the order $\mu$ part into a redefinition of $P_e$.  We can also define another viscosity  2-tensor, 
\[
\Pi_j= \Lambda_{jv}\nabla\vv{v}+\Lambda_{jj}\nabla\left(\frac{\vv{j}}{\rho}\right)\,.
\]
% which is applied to  the current field and the conductivity flux $\vv{J}_j:= \eta_{jj}\vv{j}+\eta_{jT}\nabla T$ that  drives the current. 

Given the above, from conservation of energy, one can get the new equation for entropy. The global entropy is defined by $\sigma=\sigma_i+\sigma_e$ just like the global  internal energy density is defined as $u_{\textnormal{\tiny{Vol}}}= u_{i, \textnormal{\tiny{Vol}}}+u_{e, \textnormal{\tiny{Vol}}}$. This equation is a natural generalization of the entropy evolution for magnetic field-free flow \citep{Groot84book}. Because the derivation follows directly  from the thermodynamic identity and the above equations,  we exclude the details of the calculation, which yields the following:
\bq
\ddp{\sigma}{t}+\vv{v}\cdot\nabla\sigma+\nabla\cdot \vv{J}_T=\frac{1}{T}\nabla T\cdot \vv{J}_T+\frac{1}{T}\vv{j}\cdot \vv{J}_j+\frac{1}{T}\nabla\vv{v}\!:\Pi_v+\frac{1}{T}\nabla\left(\frac{\vv{j}}{\rho}\right)\!:\Pi_j\,,
\label{entropyevol}
\eq
where we have defined the heat flux $\vv{J}_T= \eta_{Tj}\vv{j}+\eta_{TT}\nabla T$ that drives the entropy.  Here  we have added the two mirror coefficients of conductivity. The first is  the usual heat conductivity $\eta_{TT}$.  This coefficient  is  usually defined with a factor $T$, which here for simplicity and symmetry is  absorbed in the definition of the phenomenological coefficient.  Indeed, it will appear that the natural variable is $\nabla\left( {1}/{T}\right)$ and not $\nabla T$;  we  then add $T$ to  the phenomenological coefficients to compensate. The second coefficient is the cross effect conductivity $\eta_{Tj}$, which is expected to be symmetric with $\eta_{jT}$ because of the Onsager relations. 

Next we  specify the pressure and temperature. Just like before, we suppose there is a known  total internal energy density $u_{\textnormal{\tiny{Vol}}}(\rho,\sigma)$, whence the  temperature  $T$ and total pressure $P$ are given by
\[
T=\ddp{u_{\textnormal{\tiny{Vol}}}}{\sigma}\qquad \mathrm{and}\qquad  P=\frac{1}{\rho^2}\ddp{u_{\textnormal{\tiny{Vol}}}}{\rho}\,.
\]
It remains to  specify the electron pressure $P_e$ that  appears in the Ohm's law of (\ref{ohm}).  In a  more general study we might  keep the electron entropy $\sigma_e$ and define it  as in \cite{pjmK14}. In a nondissipative study, this is not a problem since the entropies are conserved. But now they evolve and it is tricky to find out which kind of dissipation varies with each entropy, since some forms of dissipation are  exchange terms, \textit{i.e.},  they  exchange entropy. Indeed, the thermodynamic study has four variables that evolve with time: the two internal energies and the two entropies, which are linked by two thermodynamic identities. The usual study takes advantages of the conservation of energy to close the system. But here we still lack an energy  equation. To compensate, we have chosen  a common  local equilibrium temperature. This is a strong hypothesis since plasmas are often not thermalized in this way.   In reality, electrons would equilibrate at one rate and ions at another, with both eventually equilibrating to a common temperature. Collisional processes could be added to account for this, but if the temperatures are initially close to each other our choice should be  a good approximation. Moreover, as noted above, this paper is already technical, and the techniques developed do  point to  the way to more complete models. Under this assumption, we suppose a known expression for  the electron Helmholtz free energy density $f_e:= u_{e, \textnormal{\tiny{Vol}}}-T\sigma_e$, which is a function of $(\rho_e, T)$,  and this  is determined for all time by our set of equations. We then can define

\[
P_e=\frac{1}{\rho_e^2}\ddp{f_e}{\rho_e}\,.
\]
If we no longer assume the local temperature equilibrium, the problem becomes harder. The global thermodynamic identity uses two temperatures, so the entropy evolution is more complex.  We save this study for future work.

\subsection{Reduced Equations and Discussion}

As before, we eliminate $\vv{j}$ and $\vv{E}$ and write the equations in terms of  $(\rho, \sigma, \vv{m}, \vv{B^*})$,  four evolution equations and four phenomenological equations.

The first evolution equation is mass conservation,
\[
\partial_t\rho+\nabla\cdot\vv{m}=0\,;
\]
the second  is momentum  conservation,
\[
\partial_t \vv{m}+\nabla\cdot(\vv{m}\otimes\vv{v})=-\nabla P-\mu \rho\frac{\chi^2}{2}\nabla\left(\frac{|\vv{j}|^2}{\rho^2}\right) +\vv{j}\times\vv{B^*}+\nabla\cdot\Pi_v\,;
\]
the third is  the magnetic field evolution equation, 
\begin{eqnarray*}
	\partial_t\vv{B^*}&=&\nabla\times
	\bigg[\vv{v}\times\vv{B^*}-\frac{\chi}{\rho}\left(\nabla\times\vv{B}\right)\times\vv{B^*}+\mu\frac{\chi}{\rho}\left(\nabla\times\vv{B}\right)\times\left(\nabla\times\vv{v}\right)
	\\
	&&\hspace{3cm}  -\vv{J}_j+\frac{1}{\rho}\nabla\cdot\Pi_j \bigg] 
	 -\frac{\chi}{\rho^2}\left(\nabla P_e \times \nabla \rho\right)\,;
\end{eqnarray*}
and the forth is  the entropy equation,
\begin{eqnarray*}
\partial_t\sigma+\nabla\cdot\left(\sigma\vv{v}\right)&=&\nabla\cdot \vv{J}_T+\frac{1}{T}\nabla T\cdot \vv{J}_T+\frac{1}{T}\vv{j}\cdot \vv{J}_j
\nonumber\\
&& \hspace{1cm} \   +\frac{1}{T}\nabla\vv{v}:\Pi_v+\frac{1}{T}\nabla\left(\frac{\vv{j}}{\rho}\right):\Pi_j\,.
\end{eqnarray*}

On the other hand, the first  two phenomenological equations determine  the  viscosities,
\[
\Pi_v= \Lambda_{vv}\nabla\vv{v}+\Lambda_{vj}\nabla\left(\frac{\vv{j}}{\rho}\right)
\qquad\mathrm{and}\qquad 
 \Pi_j= \Lambda_{jv}\nabla\vv{v}+\Lambda_{jj}\nabla\left(\frac{\vv{j}}{\rho}\right)\,;
\]
while the  second two  the conduction,
\[
 \vv{J}_j= \eta_{jj}\vv{j}+\eta_{jT}\nabla T
 \qquad\mathrm{and}\qquad  
 \vv{J}_T= \eta_{Tj}\vv{j}+\eta_{TT}\nabla T\,.
\]

Let us discuss the several dissipative effets. Some of them are usual:  the  tensor $\Lambda_{vv}$ is the usual viscosity that  gives viscous dissipation of the velocity if this velocity has spatial variation; the matrix $\eta_{jj}$ is the usual electrical resisitivity that gives the usual Ohm's law; and the  coefficient $T\eta_{TT}$ is the usual  heat conductivity that gives the usual Fourier  law. The other coefficients are less usual. The cross terms $\eta_{jT}$ and $\eta_{Tj}$ are thermo-electric coefficients, which arise from different responses of the different particles to the gradient of the temperature. In this context, this phenomenon is called thermophoresis. More precisely, $\eta_{jT}$ gives the Soret effet, while $\eta_{Tj}$ gives the Dufour effect. 

The  current viscosity  $\Lambda_{jj}$ is predominately determined by  the electron viscosity. This seems natural in the context of XMHD since the electrons do have inertia. Also, one might  think there could be  contributions of higher order in the  exchange terms, emerging from electron-ion collisions. 

According to two-fluid theory, if the electrons have viscosity, then the coefficient $\Lambda_{jj}$ will not vanish.  However,   in magnetofluid models this effect is neglected and does not appear to have been studied.    One can understand this by roughly estimating the order of magnitude of a viscosity coefficient. Dimensionally one has $[\Lambda ]=\textnormal{M}\cdot \textnormal{L}^{-1}\!\cdot \textnormal{T}^{-1}$. Choosing parameters as the mass density $nm_\alpha$,  where $\alpha\in\{ e, i\}$, one can estimate, with particle number density $n$,  the average velocity $v_\alpha$ (or equivalently the thermal energy $T$),  and the Debye length $\lambda^2_D=\frac{T}{4\pi n e^2}$, which is clearly the smallest length possible here although any length will do, we obtain
 \[
|\Lambda_{\alpha\alpha}|\sim m_\alpha n v_\alpha \lambda_D\sim n\sqrt{Tm_\alpha}\lambda_D\,.
\]
Since $n$ and $T$ are the same for both electrons and ions, one can estimate 
\[
\frac{|\Lambda_{ee}|}{|\Lambda_{ii}|}\sim\sqrt{\frac{m_e}{m_i}}=\sqrt{\mu}\,.
\]
Physically, the idea is that because electrons are much lighter,  there are more electron-electron  than ion-ion collisions, mostly because they are quicker, yet  these collisions contribute less to the change of momentum.

Alternatively, one can arrive at this result from the work of \cite{braginskii}.  Using Braginskii's coefficients $\eta_{0e}$ and $\eta_{0i}$ as estimates for the order of magnitude of $\Lambda_{ee}$ and $\Lambda_{ii}$, and assuming equal temperatures $T_e = T_i$,  one has  ${|\Lambda_{ee}|}/{|\Lambda_{ii}|} \sim \tau_e/\tau_i$, with $\tau_\alpha$ being the collision time for $\alpha \in \{e,i\}$; because $\tau_\alpha \propto \sqrt{m_\alpha}$ one reaches the same conclusion.
 
 Finally, in order of magnitude, one has $|\Lambda_{vv}|\sim|\Lambda_{ii}|\sim 1$ and $|\Lambda_{jj}|\sim|\Lambda_{ee}|\sim \sqrt{\mu}$. Numerically, this is small, yet terms of this order are retained  in the XMHD  framework. This may explain why there appears to be  no literature on this effect,  even though our estimates suggest retaining  this term as a  higher order correction of the usual formulas. We caution that our estimates are  approximate and various temperature differences for ions and electrons may also change this result. So, keeping in mind that this effect is small, we will keep it in our equations for the purposes of generality and symmetry.  Getting general brackets  is easier this  way, since one is reminded of this symmetry, which  also appears naturally in the brackets.

With regard to  the cross effects, the Onsager relations and the constraint of entropy growth assure $|\Lambda_{ie}| \sim |\Lambda_{ie}|\leq \sqrt{|\Lambda_{ii}||\Lambda_{ee}|}$. For usual cross effects, one may believe that these coefficients also are of order $\sqrt{\mu}$. Since these cross-effects can be interesting terms and also exhibit symmetries, we will keep them too. These cross-effects may be interesting since they provide a new way to mix the velocity field and the magnetic field, of higher derivative order and linear. Then, close to  equilibrium and for fast variations, while other mixing terms may disappear, these cross effect may offer new kinds of mixing. An easy and naive example will be provided in the next subsection.

\subsection{An Example  of the Viscous Cross-Effects}

In this subsection, we will brief illustration of the effect of the viscous cross-terms by showing how they  can provide  an avenue for  transferring mechanical energy into  electromagnetic energy. The model is  drastically simplified,  and intended to be  educational for gaining insight into the meaning of these new terms. 

To this end we assume, $\Omega= \R^+_y\times\R^2$, with  translational symmetry along the $z$-axis,  \textit{i.e.},    we work with a 2D-fluid. All phenomenological tensors are assumed to be constant scalars and the internal energy is chosen so that $\rho$ remains constant and uniform, just like $T$. At $y=0$ we assume there is a wall that  oscillates  in the $x$-direction with velocity $u=u_0 e^{i\nu t}$,  with $u_0$ being the amplitude, assumed small, and $\nu$ is the frequency, assumed large. We work with the viscous boundary limit, close enough to the wall and with $u_0$ small enough to neglect all nonlinear terms. We will not exhibit the huge constraints of such hypotheses.   At the beginning of this thought experiment, there is no magnetic field, and there are no outside sources, so that the electromagnetic energy is initially zero. We ask the question, upon forcing  the fluid with such a  sinusoidal mechanical input:  What will happen?

For a classical fluid,  like a plasma without the viscous cross term, the magnetic field equation states $\textnormal{d}{\vv{B^*}}\!\!/{\textnormal{d}t}=0$. Thus, the electromagnetic energy remains zero and the problem is purely mechanical. If one defines the scalar vorticity as $\omega=\hat{z}\cdot\nabla\times \vv{v}$, then the equation of motion becomes $\rho\partial_t \omega=\Lambda_{vv}\Delta\omega$. The mechanical sinosoidal input will then give rise to a sinusoidal output $\omega=\omega_0e^{ky+i\nu t}$,  where the wavenumber $k$ will respect the dispersion relation: 
\[
k=-(1\pm i)\delta^{-1}\,,  \qquad \textnormal{where}\qquad \delta=\sqrt{\frac{2\Lambda_{vv}}{\rho\nu}}\,.
\]
The quantity $\delta$ is the boundary layer thickness and $\omega_0\sim{u_0}/{\delta}$. Physically, the sinusoidal input will create oscillations in the fluid along the same direction, and this oscillation will propagate in the $y$-direction with an exponential decrease into the  boundary layer. 

Now, if we add the other viscous terms, which are linear, the situation changes. The linear terms are now, 
\begin{eqnarray}
\partial_t\omega&=&\frac{\Lambda_{vv}}{\rho}\Delta\omega-\frac{\Lambda_{vj}}{\rho}\Delta^2B 
\nonumber\\
\partial_t B^*&=&\eta_{jj}\Delta B+\frac{\Lambda_{jv}}{\rho}\Delta\omega-\frac{\Lambda_{jj}}{\rho}\Delta^2B\,,
\nonumber
\end{eqnarray}
where the components of $B^*$ and $B$ are  along the symmetry direction. Then, the sinusoidal input appears in the magnetic field equation thanks to the $\Lambda_{jv}$ term. The solutions will then become  
\[
\omega=\omega_0e^{ky+i\nu t} \qquad \mathrm{and}\qquad  B=B_0e^{ky+i\nu t}\,,
\]
and the wavenumber $k$ is now given by the dispersion relation
\[
\left(\frac{\Lambda_{vv}}{\rho}k^2-i\nu\right)\left(\eta_{jj}k^2-\frac{\Lambda_{jj}}{\rho}k^4-i\nu\left(1-\mu\frac{\chi^2}{\rho}k^2\right)\right)+\frac{\Lambda_{vj}\Lambda_{jv}}{\rho^2}k^6=0\,.
\]
Thus,  one can link $\omega_0$ and $B_0$, while $\omega_0\sim |k|u_0$.  

Evidently, the  system of  equations obtained could be  studied more deeply, the above is sufficient for our purpose:  How should one understand this thought experiment? The wall moves  sinusoidally and the  viscous boundary layer limit means the wall will drive the fluid with it. Then,  these oscillations  propagate in the $y$-direction, so the wall moves  a column of fluid. And it does so with some effectiveness, parametrized by the viscous coefficient. But which fluid is drifted? The ions or the electrons? In fact, both, but not with the same effectiveness, \textit{i.e.}, not with the same force. Thus, electrons and ions oscillate, but not at the same amplitude, thereby creating a current. If one looks at the scene from the ions' point of view, one would  see electrons oscillating. Indeed, this interface effect is creating an alternating  electric current from a mechanical input.  Thus, we have  a sort of wall-driven dynamo effect.

\section{Thermodynamic Theory of Dissipative Brackets}
\label{sec:bracket}

\subsection{From Nonequilibrium Thermodynamics to Dissipative Brackets}

While thermodynamics historically deals with equilibrium states, nonequilibrium thermodynamics is concerned with   systems close to thermal equilibrium and implements irreversible processes \citep{Groot84book, Gay-Balmaz17i, Gay-Balmaz17ii}. In developing such a theory, the first step is to write a thermodynamic identity 
\bq
\ddd \sigma=\sum_\alpha X^\alpha\ddd \zeta_\alpha\,,
\label{thermoID}
\eq
where, as before,  $\sigma$ is the entropy density and the $\zeta_\alpha$ are  densities associated with  conserved extensive properties, with $X^\alpha=\p{\sigma}/\p{\zeta_\alpha}$.  The  $\zeta_\alpha$ will eventually be used to define a  convenient  set of dynamical variables. (See section \ref{ssec:hydro}.) One then has conservation equations for all the densities, 
\[
\partial_t\zeta_\alpha+\nabla\cdot\vv{J}_\alpha=0\,, 
\]
where $\vv{J}_\alpha$ is an unknown  flux associated with $\zeta_\alpha$. Given the above, the evolution of the entropy is  determined by the equation of motion
\[
\partial_t \sigma+\nabla\cdot\vv{J}_T=\sum_\alpha \vv{J}_\alpha\cdot\nabla X^\alpha; \hspace{1.2cm} \vv{J}_T=\sum_\alpha X^\alpha\vv{J}_\alpha\,, 
\]
and $\nabla X^\alpha$ is called the affinity associated with the density and flux labeled by $\alpha$.

It remains to determine the fluxes $\vv{J}_\alpha$. Close to equilibrium, one typically assumes linear response:  
\[
\vv{J}_\alpha=\sum_\beta L_{\alpha\beta}\nabla X^\beta\,,
\]
for any $\alpha$. Up to this point, $L$ could be any tensor. But of course, physics constrains its form. Because of Onsager's relations, $L$ should be symmetric. Plus, the growth of entropy is assured if and only if $L$  has nonnegative  eigenvalues.  We make an important connection by associating dynamics generated with  a bracket with   the tensor  $L$. That is, we first prove a formal equivalence between the classical out-of-equilibrium thermodynamics and a subclass of metriplectic dynamical systems. Thus, we show  that the pseudometric nature of the dissipative bracket, usually an \textit{ad hoc} hypothesis, is the exact transcription of the well-known second law of thermodynamics and Onsager's relations through this equivalence.

It is then natural to define the phase space, a vector space of functions  on $\Omega$, that  has  the basis $\{\zeta_\alpha,\forall \alpha\}$, in which the entropy is geometrically constructed;  thus,  the form of (\ref{thermoID})  would be maintained for another choice of basis besides $\{\zeta_\alpha,\forall \alpha\}$ provided it defines a suitable set of thermodynamic variables \citep{callen}. Then, one can decompose the phase space into a part defined by the kernel of $L$ and a subspace where $L$ defines a metric. 

To see how $L$ is related to a bracket on the phase space, let us rewrite the evolution equations, at a space point $\vv{x}$ and time $t$, as follows:
\begin{eqnarray}
\partial_t\zeta_\alpha(\vv{x},t) &=&-\nabla\cdot\vv{J}_\alpha(\vv{x},t)=-\nabla\cdot\left[ L_{\alpha\beta}(\vv{x},t)\nabla \left(\ddp{\sigma}{\zeta_\beta}\right)(\vv{x},t)\right]
\nonumber\\
&=&-\int_\Omega\ddo{y}\, \delta_\Omega(\vv{x}-\vv{y})\nabla\cdot\left[ L_{\alpha\beta}(\vv{y},t)\nabla \left(\ddp{\sigma}{\zeta_\beta}\right)(\vv{y},t)\right]
\nonumber\\
&=&\int_\Omega\ddo{y}\left[\nabla\left(\delta_\Omega(\vv{x}-\vv{y})\right) L_{\alpha\beta}(\vv{y},t)\nabla \left(\ddp{\sigma}{\zeta_\beta}\right)(\vv{y},t)\right]
\nonumber\\
&=&\int_\Omega\ddo{y}\left[\nabla\left(\ddf{\zeta_\alpha(\vv{x},t)}{\zeta_\gamma(\vv{y},t)}\right) L_{\gamma\beta}(\vv{y},t)\nabla \left(\ddf{\Se(t)}{\zeta_\beta(\vv{y},t)}\right)\right], 
\label{preBkt}
\end{eqnarray}
where have been used repeated index notation for summation over $\beta$ and $\gamma$ and $\delta_\Omega(\vv{x}-\vv{y})$ is the Dirac delta function.  Proceeding from (\ref{preBkt}) one easily recognizes a  bracket, because the $\{\zeta_\alpha,\forall \alpha\}$ constitutes  a basis of the phase space.  To make this clearer, we  write the entropy evolution equation as follows:
\begin{eqnarray}
\partial_t\sigma(\vv{x},t)&=&-\nabla\cdot\vv{J}_T(\vv{x},t)+\vv{J}_\alpha\cdot\nabla\left(\ddp{\sigma}{\zeta_\alpha}\right)(\vv{x},t)
\nonumber\\
&=&-\nabla\cdot\left[\ddp{\sigma}{\zeta_\alpha}(\vv{x},t) L_{\alpha\beta}(\vv{x},t)\nabla \left(\ddp{\sigma}{\zeta_\beta}\right)(\vv{x},t)\right]
\nonumber\\
&&\hspace{2cm}+\nabla\left(\ddp{\sigma}{\zeta_\alpha}\right)(\vv{x},t) L_{\alpha\beta}(\vv{x},t)\nabla \left(\ddp{\sigma}{\zeta_\beta}\right)(\vv{x},t)
\nonumber\\
&=&-\ddp{\sigma}{\zeta_\alpha}(\vv{x},t)\nabla\cdot\left[L_{\alpha\beta}(\vv{x},t)\nabla \left(\ddp{\sigma}{\zeta_\beta}\right)(\vv{x},t)\right]\nonumber\\
&=&\int_\Omega\ddo{y}\left[\nabla\left(\ddp{\sigma}{\zeta_\alpha}(\vv{y},t)\delta_\Omega(\vv{x}-\vv{y})\right) L_{\alpha\beta}(\vv{y},t)\nabla \left(\ddp{\sigma}{\zeta_\beta}(\vv{y},t)\right)\right]
\nonumber\\
&=&\int_\Omega\ddo{y}\left[\nabla\left(\ddf{\sigma(\vv{x},t)}{\zeta_\alpha(\vv{y},t)}\right) L_{\alpha\beta}(\vv{y},t)\nabla \left(\ddf{\Se(t)}{\zeta_\beta(\vv{y},t)}\right)\right].
\nonumber
\end{eqnarray}
  
 Thus, the dynamics of out-of-equilibrium thermodynamics on the phase space can be express with a symmetric bracket. Namely, for any two functionals $f$, $g$, we define the bracket 
 \bq
 (f,g):=\frac{1}{\T}\int_\Omega\ddo{y}\, \nabla\left(\ddf{f}{\zeta_\alpha(\vv{y})}\right)\!L_{\alpha\beta}\nabla\left(\ddf{g}{\zeta_\beta(\vv{y})}\right)\,.
 \label{genBkt}
 \eq
 Here  the phenomenological tensor  $L_{\alpha\beta}$ is written with explicit subscripts $\alpha$ and $\beta$ denoting the processes, while other  tensorial indices are suppressed.   Let us remark that it is independent of the basis $\{\zeta_\alpha, \forall \alpha\}$. Indeed, the functional derivatives can be seen as  functional gradients, and both functional gradients are contracted thanks to the pseudometric $L$. We then deal with purely geometrical objects.  Similarly, if $\zeta_\alpha$ is an $a$-tensor and 
 $\zeta_\beta$ is an $b$-tensor, then  $L_{\alpha\beta}$ is an $(a+b+2)$-tensor.  Plus,  notice that thanks to the Onsager relations, the bracket is symmetric. Finally, one can write the evolution of any functional $f$ as
  \[
  \ddt{f}{t}=\T(f,\Se)\,.
  \]
 
Our construction above shows that the  dissipative brackets are completely natural for nonequilibrium thermodynamics, just like  Poisson brackets are natural for Hamiltonian dynamics.  Above, we have explicitly derived a general dissipative bracket, apparently for the first time,  from basic thermodynamic first principles.   This bracket is general and covers existing fluid-like theories of nonequilibrium thermodynamics such as that originally  given  by \citet{Morrison84b} and then  others \citep{Materassi12,Grmela97i, Edwards98}.

Consider now the  role of entropy $\Se$. It plays a role counterpart to the role of the Hamiltonian in analytical mechanics; however, of  course here, it is not a conserved quantity. On the contrary, the nonnegativity of the pseudometric $L$ assures  the entropy growth, \textit{i.e.}\  the second law of thermodynamics. We have seen in noncanonical Hamiltonian mechanics that using $\sigma$ as a variable was useful because its integral $\Se$,  being a Casimir invariant, is conserved. Another variable that  is very important, but not often  a basic dynamical variable,  is $\varepsilon$, the energy density, since it appears in the Hamiltonian that  generates  the dynamics. In thermodynamics,  $\sigma$ is no longer a natural independent variable. But $\varepsilon$, since the energy is preserved,  is a natural variable for the thermodynamic identity and then for the basis. While $\sigma$ now appears through the entropy $\Se$ that generates  the dynamics. In brief, the roles of $\varepsilon$ and $\sigma$ are interchanged. Thus, there are natural variables for the  basis of the phase space, but these are different in   the  Hamiltonian and  thermodynamic points of view. Nevertheless, one can change variables,  thanks to the thermodynamic identity, and obtain a  bracket in any complete set of phase space variables.

Finally we ask: What about the first law of thermodynamics? Using $\varepsilon$ as a basic variable in the thermodynamic framework makes it clear. Indeed, since $\delta {\Ha}/\delta {\varepsilon}$ is unity, a uniform constant,  and the other elements of the basis are independent of  $\varepsilon$, one gets for any functional $f$, 
 \[
 (f,\Ha)=0\,.
 \]
This is not a coincidence, since by construction the nonequilibrium thermodynamics conserves $\int_\Omega \zeta_\alpha$ for any $\alpha$,  and $\varepsilon$ is chosen as one of the $\zeta_\alpha$'s.  Therefore, by construction, our dissipative bracket has a strong formulation of the first law of thermodynamics.  Indeed it has all of the properties given in \cite{Morrison84,Morrison84b,Morrison86}, including bilinearity, symmetry, and degeneracy.  Coupling such a bracket with the associated noncanonical Poisson bracket gives a metriplectic dynamical system. 
 
 \subsection{Application to Hydrodynamics}
 \label{ssec:hydro}
 
 For hydrodynamics our formalism reduces to  a single fluid, without the electromagnetic effects, and the 
equations of this model are
\begin{eqnarray*}
\ddt{\rho}{t}&=&-\rho\nabla\cdot\vv{v}\\
\rho\ddt{\vv{v}}{t}&=&-\nabla P+\nabla\cdot\left(\Lambda\nabla\vv{v}\right)\\
\ddt{\sigma}{t}&=&\nabla\cdot \left(\frac{1}{T}\kappa\nabla T\right)+\frac{1}{T^2}\nabla T\cdot\kappa\nabla T+\frac{1}{T}\nabla\vv{v}:\Lambda\nabla\vv{v},\,
\end{eqnarray*}
where $\Lambda$ is the viscosity, a 4-tensor,  and $\kappa$ the heat conductivity, in general a 2-tensor.

 We now apply our new formulation to the fluid, whose variables are $(\varepsilon, \rho, \vv{m})$. Indeed, these variables are independent, they specify the state of the fluid,  and they are conserved densities;  respectively,  the total energy $\Ha$, the global momentum $\boldsymbol{\mathcal{P}}$,  and the total mass $\mathcal{M}$ are constants of motion: 
\[
\Ha=\int_\Omega\varepsilon\,,\quad  \boldsymbol{\mathcal{P}}=\int_\Omega\vv{m}\,, \qquad  \mathrm{and}\qquad
\mathcal{M}=\int_\Omega\rho\,.
 \]
 Our construction guarantees that these quantities will remain constant. 
 If $u$ is the specific  internal energy, the local energy density is
 \[
\varepsilon=\frac{|\vv{m}|^2}{2\rho} +  \rho u(\rho, s)\,,
\]
where $s$ is the specific entropy and $\sigma=\rho s$. 
 The thermodynamic identity reads $\ddd u=T\ddd s+ {P}\ddd \rho/\rho^2$, which upon  changing variables gives
\[
T\ddd\sigma=\ddd\varepsilon-\vv{v}\cdot\ddd\vv{m}-g\ddd\rho\,,
\]
where $g$ is a modified specific Gibbs free energy, namely $g:= u-Ts+{P}/{\rho}-|{\vv{v}|^2}/{2}$. Its differential is then $\ddd g=-s\ddd T+ \ddd P/\rho-\vv{v}\cdot\ddd\vv{v}$, so that $g(T,P,\vv{v})$ is an extensive quantity  with intensive arguments and, consequently,   vanishes.  Let us remark that in this paper we do not consider chemical reactions or particle creation/annihilation --  for  such cases this free energy would not vanish. Finally, the phase space for thermodynamics is smaller than that for the Hamiltonian case because $\rho$ does not appear in the thermodynamic identity and may be ignored. The thermodynamic variables then are $(\varepsilon, \vv{m})$ and the thermodynamic identity is
\[
T\ddd\sigma=\ddd\varepsilon-\vv{v}\cdot\ddd\vv{m}\,.
\]
 
From this thermodynamic identity, one can see that there will be two irreversible responses, linked to $\varepsilon$ and $\vv{m}$, that could be expressed with the affinities $\nabla T$ and $\nabla\vv{v}$. These dissipation processes are,  respectively,  the heat conduction and the viscosity. We are set to proceed, but  there are two  complications. First, the natural affinity associated with $\varepsilon$ is $\nabla\left({1}/{T}\right)$, but this choice implies  some factors of $T$ will appear. Second, the natural affinity of  $\vv{m}$ is $-\nabla\left({\vv{v}}/{T}\right)$, but this  may create cross effects. Since we know  because of space-parity symmetry no such cross effects exist between the affinities $\nabla T$ and $\nabla\vv{v}$, we must destroy them using nondiagonal terms of the tensor $L$. This is a strange constraint that  is certainly linked with the choice of the basis. 
 
To get the expression for  $L$, one can look at the various fluxes and compare them with the usual notations \citep{Groot84book}. For $\vv{m}$, the flux is the opposite of the usual viscous tensor, \textit{viz.},
\[
\Pi_v=L_{mm}\nabla\left(\frac{\vv{v}}{T}\right)-L_{m\varepsilon}\nabla\left(\frac{1}{T}\right)=\Lambda\nabla\vv{v}\,,
\]
 whence $L_{mm}=T\Lambda$ and $L_{m\varepsilon}=T\Lambda\vv{v}$.   On the other hand, the energy flux is
 \[
 \vv{J}_\varepsilon=-L_{\varepsilon m}\nabla\left(\frac{\vv{v}}{T}\right)+L_{\varepsilon \varepsilon}\nabla\left(\frac{1}{T}\right)=\vv{v}\cdot\Pi_v+\kappa\nabla T\,,
 \]
whence   $L_{\varepsilon m}=T\vv{v}\cdot\Lambda$ and $L_{\varepsilon \varepsilon}=T^2\kappa+\vv{v}\cdot\Lambda\vv{v}$.  Thus,    $L$ is effectively symmetric  and it is easy to check that we have the correct  heat flux,
\[
\vv{J}_T=\frac{1}{T}\vv{J}_\varepsilon+\vv{v}\cdot\Pi_v=\frac{1}{T}\kappa\nabla T\,.
\]

Having proceeded in this  systematic way, the bracket on any functionals $f$ and $g$ is  immediate:
	\begin{eqnarray}
		 (f,g)&=&\frac{1}{\T}\int_{\Omega}\ddo{y}\, \Bigg(\nabla\left(\ddf{f}{\varepsilon(\vv{y})}\right)\cdot \Bigg[ \left(T^2\kappa+\vv{v}\cdot\Lambda\vv{v}\right)\nabla\left(\ddf{g}{\varepsilon(\vv{y})}\right)
		+T\vv{v}\cdot\Lambda\nabla\left(\ddf{g}{\vv{m}(\vv{y})}\right)\Bigg]
		\nonumber\\
		&&\hspace{1cm}+\nabla\left(\ddf{f}{\vv{m}(\vv{y})}\right):\left[T\Lambda\vv{v}\otimes\nabla\left(\ddf{g}{\varepsilon(\vv{y})}\right)+T\Lambda\nabla\left(\ddf{g}{\vv{m}(\vv{y})}\right)\right]\Bigg).
	\end{eqnarray}
We know this bracket is a metriplectic bracket that preserves the desired quantity, it  is symmetric,  and it is positive. 
	
	What about the known dissipative bracket of hydrodynamics given by  \citet{Morrison84b}?  To compare we transform  back to the  more usual fluid dynamical variables of fluid mechanics, $(\varepsilon, \vv{m}, \rho)\longrightarrow (\sigma, \vv{m}, \rho)$. Via the chain rule the  functional derivatives satisfy
	\[
\ddf{f}{\varepsilon}\longrightarrow \ddp{\sigma}{\varepsilon}\ddf{f}{\sigma}=\frac{1}{T}\ddf{f}{\sigma}
\quad \mathrm{and}\quad
 \ddf{f}{\vv{m}}\longrightarrow \ddf{f}{\vv{m}}+\ddp{\sigma}{\vv{m}}\ddf{f}{\sigma}=\ddf{f}{\vv{m}}-\frac{\vv{v}}{T}\ddf{f}{\sigma}\,.
\]
Using
\[
\nabla \left(\ddf{f}{\vv{m}}-\frac{\vv{v}}{T}\ddf{f}{\sigma}\right)=\nabla \left(\ddf{f}{\vv{m}}\right)-\frac{1}{T}\ddf{f}{\sigma}\nabla\vv{v}-\nabla\left(\frac{1}{T}\ddf{f}{\sigma}\right)\otimes\vv{v}\,,
\]
it is seen that  the last term compensates the cross terms,  simplifying  the heat part of the bracket, yielding, 
	\begin{eqnarray}
	\hspace{-2cm}(f,g)&=& \int_\Omega \ddo{y}\, \frac{T}{\T}\Bigg[T\nabla\left(\frac{1}{T}\frac{\delta f}{\delta\sigma(\vv{y})}\right)\cdot \kappa\nabla\left(\frac{1}{T}\frac{\delta g}{\delta\sigma(\vv{y})}\right)
	\\
	&& +\left(\nabla\left(\frac{\delta f}{\delta \vv{m}(\vv{y})}\right)-\frac{1}{T}\frac{\delta f}{\delta\sigma(\vv{y})}\nabla\vv{v}\right):\Lambda \left(\nabla\left(\frac{\delta g}{\delta \vv{m}(\vv{y})}\right)-\frac{1}{T}\frac{\delta g}{\delta\sigma(\vv{y})}\nabla\vv{v}\right)\Bigg]\,,
		\nonumber
	\end{eqnarray}
the bracket given by  \cite{Morrison84b}. 

To summarize, in this section we have developed a systematic way to construct dissipative brackets,  and we showed that this method reproduces the known bracket for  hydrodynamics.  In the next section we will use the method to derive a  bracket for XMHD.  There we  will change some notation, \textit{i.e.},   we write for the 2-tensor $\eta_{TT}:=\kappa/T$ and the 4-tensor $\Lambda_{vv}:=\Lambda$.

\section{Derivation of the Brackets of Dissipative Extended Magnetohydrodynamics}
 \label{sec:detbracket}
 
\subsection{Thermodynamics}

We now return to XMHD.  Recall, for this theory the  energy density $\varepsilon$  is given by the expression of (\ref{xmhdE}) with  $\vv{B^*}$  given by (\ref{Bstar}).  For dissipative XMHD, this quantity  needs to be incorporated into the theory via an appropriate choice of a magnetic conserved quantity. 
Then,  we can choose the associated magnetic variable and use directly the general bracket theory  derived in section \ref{sec:bracket}.

Since  $\varepsilon$ now depends on the magnetic field, the thermodynamic identity will be modified accordingly.  First, note that the global momentum $\boldsymbol{\mathcal{P}}=\int_\Omega\vv{m}$ is conserved, consistent with the Galilean symmetry as realized by the noncanonical bracket \citep{pjm82}. Because of quasineutrality the local momentum $\vv{m}$ has no magnetic (vector potential) piece (as explained in \cite{Charidakos14}), and so remains equal to $\rho\vv{v}$.  Consequently, $\boldsymbol{\mathcal{P}}$ will not participate in the magnetic part of the dissipation.  One can check easily from the equations of  dissipative XMHD of section \ref{sec:DXMHD} that  the integrated drifted magnetic field $\int_\Omega\vv{B^*}$ is preserved. Thus, we will use the coordinates $(\varepsilon, \rho, \vv{m}, \vv{B^*})$ and express the thermodynamic identity in terms of these variables. 

 One may ask why $\int_\Omega\vv{B^*}$ should be conserved, which unlike  other conserved quantities does not come from deeper insight such as Galilean invariance.  To explain this,  let us first digress for a moment and address why $\int_\Omega\vv{B}$ is conserved for ordinary magnetohydrodynamics,  a fact  pointed out as early as  \citet{pjm82} that did not seem to be well known.  To interpret this conservation law,  consider the postulate of  conservation of the magnetic flux of any surface moving with the  velocity field $\vv{v}$. First, remember that the magnetic field is a pseudovector (vector density), which is more naturally re-expressed as a 2-form. In local coordinates, this 2-form, denoted $\mathfrak{B}$, can be written as $\mathfrak{B}_{ab}=\varepsilon_{abc}\vv{B}_c$,  where $\varepsilon_{abc}$ is  the Levi-Civita tensor. Thus, the magnetic flux through a surface $\Sigma$ with unit normal  $\vv{n}$ is $\int_\Sigma \vv{B}\cdot\vv{n}=\int_\Sigma i^*\mathfrak{B}$,  where $i^*$ is  the pull back of the inclusion $\Sigma\subset\Omega$ that chooses the associated coordinate of the 2-form $\mathfrak{B}$. Given the assumption that the flux through a surface is preserved when advected by  $\vv{v}$, not forgetting  that the surface $\Sigma$ moves following this vector field, we get the equation 
\begin{equation*}
\ddt{\ }{t} \int_\Sigma i^*\mathfrak{B} =\int_{\Sigma}i^*\left(\partial_t\mathfrak{B}+\pounds_\vv{v}\mathfrak{B}\right)=0\,,
\end{equation*}
where $\pounds_\vv{v}$ is the Lie derivative generated by  the vector field $\vv{v}$. So, our postulate is equivalent to the local Lie dragging  equation 
\bq
\label{advected magnetic field}
	\partial_t\mathfrak{B}+\pounds_\vv{v}\mathfrak{B}=0\,. 
\eq
Now, choose some fixed direction to evaluate the magnetic field, that is, a 1-form $\theta$. We can restrict ourselves to closed 1-forms, \textit{i.e.},   $\textnormal{d}\theta=0$. Remembering that the magnetic field being  divergence-free means  $\textnormal{d}\mathfrak{B}=0$, and using the Cartan formula and   Stokes theorem, we have the identity 
\bq
\label{Lie alpha}
	\int_\Omega \mathfrak{B}\wedge\pounds_{\vv{v}}\theta= \int_\Omega \mathfrak{B}\wedge\textnormal{d}i_\vv{v}\theta=\int_\Omega \textnormal{d}(\mathfrak{B}\wedge i_\vv{v}\theta)-\int_\Omega (\textnormal{d}\mathfrak{B})\wedge i_\vv{v}\theta=0\,,
\eq
where $i_\vv{v}$ is the interior product by $\vv{v}$. The integral of the $\theta$-component of the magnetic field is $\int_\Omega \theta(\vv{B})=\int_\Omega \mathfrak{B}\wedge \theta$. One can interpret this as a global flux, the sum of the fluxes through the local surfaces normal to $\theta$. Using that the vector field $\vv{v}$ preserves $\Omega$,  then that $\theta$ is fixed and satisfies  equation (\ref{Lie alpha}), the evolution of this property is 
\bqy
\ddt{\ }{t} \int_\Omega \theta(\vv{B}) &=&\int_{\Omega}\left(\partial_t(\mathfrak{B}\wedge\theta)+\pounds_{\vv{v}}(\mathfrak{B}\wedge\theta)\right)
\nonumber\\
&=&\int_{\Omega}\left(\partial_t\mathfrak{B}+\pounds_{\vv{v}}\mathfrak{B}\right)\wedge\theta=0\,.
\nonumber
\eqy
Finally, saying that this is true for any $\theta$ is just saying that $\int_\Omega \vv{B}$ is preserved, which is exactly what we wanted. Note, here the vector field $\vv{v}$ could have been any vector field, and need not be the physical velocity field, since we only used that it preserves the domain $\Omega$.

The flux conservation assumption is central  for magnetohydrodynamics with the velocity field being the  Lie dragging field $\vv{v}$, which is  Alfven's  well-known  \textit{frozen-in} property (\textit{e.g.},  \cite{Kampen67book}). Nevertheless, this is no longer true for XMHD, since neither $\vv{B}$ nor $\vv{B}^*$ are advected  by $\vv{v}$.   Yet, the conservation of $\int_\Omega\vv{B}^*$ can be seen by  using modified velocity fields. Indeed, according to \citet{DAvignon16} (and with their notation), the drifted magnetic field can be decomposed into the following form: 
\begin{equation}
	\vv{B}^*=\frac{\beta_-\vv{B}_+-\beta_+\vv{B}_-}{\beta_--\beta_+}\,, 
\end{equation}
where $\beta_\pm$ are scalar constants and $\vv{B}_\pm$ are modified `magnetic fields' that  satisfy  equation (\ref{advected magnetic field}) with modified velocities $\vv{v}_\pm$. Thus, the integrals of $\vv{B}_\pm$ are preserved, and by linearity, the integral $\int_\Omega\vv{B}^*$ is  too.

Given that we have settled on our magnetic conserved quantity, let us proceed with obtaining the dissipative bracket.  To this end we will need that the variation of $\vv{B^*}$ can be expressed as
\[
\delta\vv{B^*}=\delta\vv{B}+\mu\chi^2\nabla\times\left(\frac{1}{\rho}\nabla\times\delta\vv{B}\right)-\mu\chi^2\nabla\times\left(\frac{\vv{j}}{\rho^2}\,\delta\rho\right)\,.
\]
 Remembering that variations will be integrated,  we can directly write,  using  integration by part, the following: 
	\begin{eqnarray*}
		\vv{B}\cdot\delta\vv{B^*}&=&\vv{B}\cdot\delta\vv{B}+\mu\chi^2\vv{B}\cdot\nabla\times\left(\frac{1}{\rho}\nabla\times\delta\vv{B}\right)-\mu\chi^2\vv{B}\cdot\nabla\times\left(\frac{\vv{j}}{\rho^2}\delta\rho\right)
		\\
		&=&\vv{B^*}\cdot\delta\vv{B}-\mu\chi^2\left(\frac{|\vv{j}|}{\rho}\right)^2\delta\rho\,,
	\end{eqnarray*}
where the second equality is modulo  a total divergence, and similarly, since here differentials and variations are the same,
\bq
\ddd\left(\frac{\vv{B}\cdot\vv{B^*}}{2}\right)=\vv{B}\cdot\ddd\vv{B^*} 
+ \mu\chi^2\left(\frac{|\vv{j}|}{\rho}\right)^2\ddd\rho\,.
\label{bstarID}
\eq

Using (\ref{bstarID}) and (\ref{xmhdE}) we now can write the following thermodynamic identity: 
\[
T\ddd\sigma=\ddd\varepsilon-\vv{v}\cdot\ddd\vv{m}-\vv{B}\cdot\ddd\vv{B^*}-g^*\ddd\rho\,,
\]
where $g^*:= u-Ts+  {P}/{\rho}- {|\vv{v}|^2}/{2}+\mu\chi^2\left({|\vv{j}|}/{\rho}\right)^2$ is a modified specific Gibbs free energy. Thus, its natural variables are $g^*(T,P,\vv{v},{\vv{j}}/{\rho})$ or equivalently $g^*(T,P,\vv{v}_i,\vv{v}_e)$.  An   extensivity/intensity argument like that of section \ref{ssec:hydro} shows again $g^*=0$.  Thus,  the thermodynamic identity is given by
\bq
T\ddd\sigma=\ddd\varepsilon-\vv{v}\cdot\ddd\vv{m}-\vv{B}\cdot\ddd\vv{B^*}\,.
\label{xmhdTD}
\eq
Neglecting the factors of $T$ discussed in section \ref{ssec:hydro}, which we will return to, we would conclude that  the response to $\vv{m}$  is $\nabla\vv{v}$, the gradient of the thermodynamic dual and, similarly,   the response to $\vv{B^*}$ would be  $\nabla\vv{B}$.    One might  think that this latter response is a 2-tensor; actually, it is not because $\vv{B}$, as noted above,  is not a vector but a pseudovector or more naturally a 2-form, with the constraint $\nabla\cdot\vv{B}=0$. Then, it is natural to think that this constraint will reduce the size of the response. Indeed, we will see that the response will be $\nabla\times\vv{B}=\vv{j}$, which is a 1-tensor, and a special case of the global 2-tensor, if we re-write the phenomenological tensor. 

There is yet another complication. From the thermodynamic identity (\ref{xmhdTD}), we easily see that the response to $\varepsilon$ will create heat conduction, that the response to 
$\vv{B}^*$ will create electrical resistivity, and cross terms can, of course,  easily be added.  Also we see that the response to $\vv{m}$ will create velocity viscosity, but what about current viscosity? In fact, in the energy expression there is  a term with $\vv{B}$, the magnetic energy, and so  conduction, and a term in $\vv{j}$, the inertia of electrons,  and so the current viscosity.   So, what happened? We reduced variables to eliminate $\vv{j}$,  thanks to Ampere's law.  In  this way we hid this effect. How should we manage this situation? We will add a new affinity, one that  will create this current viscosity. The original energy had a term with ${|\vv{j}|^2}/{\rho}$,  which would create an affinity $\nabla\left({\vv{j}}/{\rho}\right)$, which is exactly what appears in the equations. 
	Thus,  we are finally let to the following list of  affinities:
$$
\nabla\left(\ddf{\Se}{\varepsilon}\right); \hspace{1.2cm} \nabla\left(\ddf{\Se}{\vv{m}}\right); \hspace{1.2cm} \nabla\times\left(\ddf{\Se}{\vv{B^*}}\right); \hspace{1.2cm} \nabla\left(\frac{1}{\rho}\nabla\times\left(\ddf{\Se}{\vv{B^*}}\right)\right)\,.
$$

Now, let us return to the problem involving $\nabla T$ encountered  in the hydrodynamic case.  Recall, the natural variables led to  the bad effect of  generating a  cross term with $\nabla T$ that must be compensated through cross terms. But our new affinity is of second order and so  will create cross terms of  second order in $T$. To compensate for these bad terms, we have to have  a new affinity of second order in $T$.  Upon examination of these cross terms, we see that the needed affinity will be 
	\[
	\nabla\left(\frac{1}{\rho}\vv{B}\times\nabla\left(\ddf{\Se}{\varepsilon}\right)\right)\,.
	\]
	One could consider a general second derivative of the temperature and modify the phenomenological tensor as a  consequence, but for simplicity we will directly use the easiest affinity.
	
	\subsection{Determination of the Phenomenological Tensor}
	
	At this point, there are two paths to get the dissipative brackets of the theory. First, in  complete analogy with the hydrodynamic case studied in section \ref{sec:bracket}, we can  identify the various coefficients of the phenomenological tensor from the known flux expressions derived in section \ref{sec:DXMHD}. Then, we have an expression for the dissipative brackets that  can easily be changed back to  the usual set of variables. Alternatively,  we can  first change variables of  the affinity expressions and then identify the phenomenological tensors through the known equations of the model. The advantage of this second method is that in the dynamical variables, there are  no tricky cross-effects due to  the gradient of temperature.  For this reason we  will follow the second method, buttressed  by  our experience with the hydrodynamic example,  and find it to be an easier calculation. Both calculations lead to the same conclusion.
	
	With the change of variables $(\varepsilon, \rho, \vv{m},\vv{B^*})\longrightarrow (\sigma, \rho, \vv{m},\vv{B^*})$, the  functional derivatives  of any functional $f$ become
	\[
	\ddf{f}{\varepsilon}\longrightarrow \frac{1}{T}\ddf{f}{\sigma};\hspace{1.1cm} \ddf{f}{\vv{m}}\longrightarrow \ddf{f}{\vv{m}}-\frac{\vv{v}}{T}\ddf{f}{\sigma}; \hspace{1.1cm} \ddf{f}{\vv{B^*}}\longrightarrow \ddf{f}{\vv{B^*}}-\frac{\vv{B}}{T}\ddf{f}{\sigma}\,.
	\]
The interesting affinities in terms of the new variables will change and,  just like for the hydrodynamic case, we develop them with the temperature away from the derivatives. Developed in this  way, we directly see the cross effect that we want to vanish. 

Now we  list the terms for any functional $f$. First, we have the $\varepsilon$ and $\vv{m}$ responses,  which are the same as those  for hydrodynamics, 
\bq
\nabla\left(\ddf{f}{\varepsilon}\right)\longrightarrow \nabla\left(\frac{1}{T}\frac{\delta f}{\delta\sigma(\vv{y})}\right)
\label{gradep}
\eq
 and
 \bq
 \nabla\left(\ddf{f}{\vv{m}}\right)\longrightarrow \nabla \left(\ddf{f}{\vv{m}}\right)-\frac{1}{T}\ddf{f}{\sigma}\nabla\vv{v}-\nabla\left(\frac{1}{T}\ddf{f}{\sigma}\right)\otimes\vv{v}\,.
\label{gradm}
\eq
	Similarly, the resistivity will arise from the magnetic response, 
	\begin{eqnarray}
		\nabla\times\left(\ddf{f}{\vv{B^*}}\right)\longrightarrow 
		&\nabla \times\left(\ddf{f}{\vv{B^*}}\right)-\frac{1}{T}\ddf{f}{\sigma}\nabla\times\vv{B} 
		-\nabla\left(\frac{1}{T}\ddf{f}{\sigma}\right)\times\vv{B}
		\nonumber\\
		&=\nabla \times\left(\ddf{f}{\vv{B^*}}\right)-\frac{1}{T}\ddf{f}{\sigma}\, \vv{j}+\vv{B}\times\nabla\left(\frac{1}{T}\ddf{f}{\sigma}\right)\,.
		\label{MR}
	\end{eqnarray}
Using (\ref{MR}) for  the new affinity, the current viscosity will change into 
	\begin{eqnarray}
		\hspace{-1cm}\nabla\left(\frac{1}{\rho}\nabla\times\left(\ddf{f}{\vv{B^*}}\right)\right)\longrightarrow &\nabla \left[\frac{1}{\rho}\nabla \times\left(\ddf{f}{\vv{B^*}}\right)-\frac{1}{T}\ddf{f}{\sigma}\frac{\vv{j}}{\rho}+\frac{1}{\rho}\vv{B}\times\nabla\left(\frac{1}{T}\ddf{f}{\sigma}\right)\right]
		\nonumber\\
		&=\nabla \left(\frac{1}{\rho}\nabla \times\left(\ddf{f}{\vv{B^*}}\right)\right)-\frac{1}{T}\ddf{f}{\sigma}\nabla \left(\frac{\vv{j}}{\rho}\right)
		\label{gradB}\\
		&\hspace{.7cm}-\nabla\left( \frac{1}{T}\ddf{f}{\sigma}\right)\otimes\frac{\vv{j}}{\rho}+\nabla\left(\frac{1}{\rho}\vv{B}\times\nabla\left(\frac{1}{T}\ddf{f}{\sigma}\right)\right)\,.
		\nonumber
	\end{eqnarray}
Lastly, the term with the second derivative of the temperature becomes 
	\bq
	\nabla\left(\frac{1}{\rho}\vv{B}\times\nabla\left(\ddf{f}{\varepsilon}\right)\right) \longrightarrow \nabla\left(\frac{1}{\rho}\vv{B}\times\nabla\left(\frac{1}{T}\ddf{f}{\sigma}\right)\right)\,.
\label{gradBep}
\eq

From our knowledge of the form of the brackets in the thermodynamic variables and the determination of the affinities in the  dynamical coordinates  $(\sigma, \rho, \vv{m},\vv{B^*})$, we can re-write the general bracket. In particular, upon replacing the densities of the general bracket of (\ref{genBkt}) by expressions   (\ref{gradep}) -- (\ref{gradBep}) above,  a bracket with many terms is generated.  For efficiency, we will directly develop the temperature terms from the different affinities and add a $T$ factor when it makes things easier. We denote by $\widehat{L}$ the phenomenological tensor in these new coordinates, with indices denoting the processes as explained in table \ref{table} while, as before,  we suppress tensorial indices that should be clear from context. Thus, the bracket has the following form:  
	\begin{eqnarray*}
	(f,g)= \int_\Omega \frac{T}{\T}\, 
	R_\alpha(f)\, \widehat{L}_{\alpha\beta}\, R_\beta(g)\,, 
	\end{eqnarray*}
where $\alpha$ and $\beta$ are summed over the process index set $\{T, v, B, j, T2\}$ and the responses $R_\alpha$ are explicitly given in table \ref{table}. These responses may be 1- or 2-tensors and are contracted with the phenomenological tensors that  are of appropriate rank; \textit{e.g.},   if $R_\alpha$ is a $a$-tensor and $R_\beta$ is a $b$-tensor, then $\widehat{L}_{\alpha\beta}$ is a $(a+b)$-tensor.
	
\begin{table}
\centering
\begin{tabular}{|c|c|c|c|c|}
\hline
Response  to $f$ & Explicit formula & Geometrical type & Index for $\widehat{L}$ \\ \hline
$R_T(f)$  & $\nabla\left(\frac{1}{T}\frac{\delta f}{\delta\sigma}\right)$ & 1-tensor & $T$ \\ \hline
$R_v(f)$  & $\nabla \left(\ddf{f}{\vv{m}}\right)-\frac{1}{T}\ddf{f}{\sigma}\nabla\vv{v}$   &  2-tensor & $v$ \\ \hline
$R_B(f)$ &  $\nabla \times\left(\ddf{f}{\vv{B^*}}\right)-\frac{1}{T}\ddf{f}{\sigma}\vv{j}$  & 1-tensor & $B$ \\ \hline
$R_j(f)$ & $\nabla \left[\frac{1}{\rho}\nabla \times\left(\ddf{f}{\vv{B^*}}\right)\right]-\frac{1}{T}\ddf{f}{\sigma}\nabla\left( \frac{\vv{j}}{\rho}\right)$ & 2-tensor & $j$ \\ \hline
$R_{T2}(f)$ & $\nabla\left(\frac{1}{\rho}\vv{B}\times\nabla\left(\frac{1}{T}\ddf{f}{\sigma}\right)\right)$ & 2-tensor & $T2$\\ \hline  
\end{tabular}
\caption[Table caption text]{The several kinds of thermodynamic responses, including their tensor character, for dissipative extended magnetohydrodynamics.}
\label{table}
\end{table}

Now, it only remains to find the phenomenological coefficients. For this, we write the various equations of XMHD and identify the terms.  We remark that thanks to the  Onsager relations,    the phenomenological tensor is symmetric  and when we determine a term we automatically know its dual term. We could use this to shorten calculations, but for completeness  we will write out all the terms and discover these symmetries. Proceeding, we see the  momentum equation gets the dissipative term
		\begin{eqnarray*}
		(\vv{m},\Se)\T&=&\nabla\cdot\bigg[\frac{1}{T}\widehat{L}_{vT}\nabla T+ \widehat{L}_{vv}\nabla\vv{v}+\widehat{L}_{vv}\vv{j}
		%\\
		%&&%\hspace{2cm}
		+\widehat{L}_{vj}\nabla\left( \frac{\vv{j}}{\rho}\right) 
		+\widehat{L}_{vT2}\nabla\left(\frac{1}{\rho T}\vv{B}\times\nabla T \right)\bigg]
		\\
		&=&\nabla\cdot\left[\Lambda_{vv}\nabla\vv{v}+\Lambda_{vj}\nabla\left(\frac{\vv{j}}{\rho}\right)\right].
	\end{eqnarray*}
Thus, we identify the nonvanishing phenomenological coefficients $\widehat{L}_{vv}=\Lambda_{vv}$ and $\widehat{L}_{vj}=\Lambda_{vj}$. 
	Next, for the magnetic field, we get
	\begin{eqnarray*} 
	 (\vv{B^*},\Se)\T&=&-\nabla\times\Bigg[\frac{1}{T}\widehat{L}_{BT}\nabla T+ \widehat{L}_{Bv}\nabla\vv{v}+\widehat{L}_{BB}\vv{j}+\widehat{L}_{Bj}\nabla\left( \frac{\vv{j}}{\rho}\right)\\
		&&\hspace{5cm}+\widehat{L}_{BT2}\nabla\left(\frac{1}{\rho T}\vv{B}\times\nabla T \right)\Bigg]\\
	&&\hspace{1.5cm} +\nabla\times\Bigg(\frac{1}{\rho}\nabla\cdot\Bigg[\frac{1}{T}\widehat{L}_{jT}\nabla T+ \widehat{L}_{jv}\nabla\vv{v}+\widehat{L}_{jB}\vv{j}+\widehat{L}_{jj}\nabla\left( \frac{\vv{j}}{\rho}\right)\\
	&&\hspace{5cm}+\widehat{L}_{jT2}\nabla\left(\frac{1}{\rho T}\vv{B}\times\nabla T \right)\Bigg]\Bigg)\\
	&=&\nabla\times\left[-\left(\eta_{jj}\vv{j}+\eta_{jT}\nabla T\right)+\frac{1}{\rho}\nabla\cdot \left(\Lambda_{jv}\nabla\vv{v}+\Lambda_{jj}\nabla\left(\frac{\vv{j}}{\rho}\right)\right)\right].
	\end{eqnarray*}
Thus, the only nonvanishing coefficients are $\widehat{L}_{BT}= T\eta_{jT}$, $\widehat{L}_{BB}= \eta_{jj}$, $\widehat{L}_{jv}= \Lambda_{jv}$,  and $\widehat{L}_{jj}= \Lambda_{jj}$.
	Finally, using the previous results, the entropy equation yields
	\begin{eqnarray*} 
(\sigma,\Se)\T&=&\nabla\Bigg[\frac{1}{T^2}\widehat{L}_{TT}\nabla T+ \frac{1}{T}\widehat{L}_{Tv}\nabla\vv{v}+\frac{1}{T}\widehat{L}_{TB}\vv{j}+\frac{1}{T}\widehat{L}_{Tj}\nabla\left( \frac{\vv{j}}{\rho}\right)\\
		&&\hspace{5cm}+\frac{1}{T}\widehat{L}_{TT2}\nabla\left(\frac{1}{\rho T}\vv{B}\times\nabla T \right)\Bigg]\\
	&&\hspace{.75cm} +\frac{1}{T^2}\nabla T\cdot \Bigg[\frac{1}{T}\widehat{L}_{TT}\nabla T+ \widehat{L}_{Tv}\nabla\vv{v}+\widehat{L}_{TB}\vv{j}+\widehat{L}_{Tj}\nabla\left( \frac{\vv{j}}{\rho}\right)\\
	&&\hspace{5cm}+\widehat{L}_{TT2}\nabla\left(\frac{1}{\rho T}\vv{B}\times\nabla T \right)\Bigg]\\
 	&&\hspace{.75cm} +\frac{1}{T}\nabla v :\Pi_v+\frac{1}{T}\nabla \left(\frac{\vv{j}}{\rho}\right) :\Pi_v+\frac{1}{T}\vv{j}\cdot\vv{J}_j\\
 	&&\hspace{.75cm} +\frac{1}{T}\nabla \cdot\Bigg(\frac{\vv{B}}{\rho}\times \nabla\cdot\Bigg[\frac{1}{T}\widehat{L}_{T2T}\nabla T+ \widehat{L}_{T2v}\nabla\vv{v}+\widehat{L}_{T2B}\vv{j}+\widehat{L}_{T2j}\nabla\left( \frac{\vv{j}}{\rho}\right)\\
 	&&\hspace{5cm}+\widehat{L}_{T2T2}\nabla\left(\frac{1}{\rho T}\vv{B}\times\nabla T \right)\Bigg]\Bigg)\\
	&=&\nabla\cdot \left(\eta_{Tj}\vv{j}+\eta_{TT}\nabla T\right)+\frac{1}{T}\nabla T\cdot \left(\eta_{Tj}\vv{j}+\eta_{TT}\nabla T\right)+\frac{1}{T}\vv{j}\cdot \vv{J}_j\\
	&&\hspace{5cm}+\frac{1}{T}\nabla\vv{v}:\Pi_v+\frac{1}{T}\nabla\left(\frac{\vv{j}}{\rho}\right):\Pi_j\,.
	\end{eqnarray*}
	Then, the nonvanishing coefficients are $\widehat{L}_{TT}= T^2\eta_{TT}$ and ${L}_{TB}= T\eta_{Tj}$. Observe from the above that  indeed  the Onsager relations hold. 
	
	From the  tensor $\widehat{L}$, one could change coordinates back and obtain the tensor $L$. This is an easy computation, but not of our interest here. Similarly, one could have computed the tensor $L$ before changing coordinates to get the tensor $\widehat{L}$.
	
	\subsection{Metriplectic Framework}
	
	We are now able to write the dissipative part of the  metriplectic bracket for  XMHD. From our calculations, we have found the following bracket for any functionals $f$ and $g$:	
\begin{eqnarray}
 (f,g)&=&  \int_\Omega  \ddo{y}\, \frac{T}{\T}\Bigg[\nabla\left(\frac{1}{T}\frac{\delta f}{\delta\sigma(\vv{y})}\right)\cdot \Bigg\{T^2\eta_{TT} \nabla\left(\frac{1}{T}\frac{\delta g}{\delta\sigma(\vv{y})}\right)
\nonumber\\
	&&\hspace{3.5cm}+ T\eta_{Tj} \left(\nabla\times\left(\frac{\delta g}{\delta \vv{B^*}(\vv{y})}\right)-\frac{1}{T}\frac{\delta g}{\delta\sigma(\vv{y})}\vv{j}\right)\Bigg\}
	\nonumber\\
	&& +\left(\nabla\times\left(\frac{\delta f}{\delta \vv{B^*}(\vv{y})}\right)-\frac{1}{T}\frac{\delta f}{\delta\sigma(\vv{y})}\vv{j}\right)\cdot \Bigg\{ T\eta_{jT} \nabla\left(\frac{1}{T}\frac{\delta g}{\delta\sigma(\vv{y})}\right)
	\nonumber\\
	&&\hspace{3.5cm}+\eta_{jj} \left(\nabla\times\left(\frac{\delta g}{\delta \vv{B^*}(\vv{y})}\right)-\frac{1}{T}\frac{\delta g}{\delta\sigma(\vv{y})}\vv{j}\right)\Bigg\}
	\nonumber\\ 
	&& +\left(\nabla\left(\frac{\delta f}{\delta \vv{m}(\vv{y})}\right)-\frac{1}{T}\frac{\delta f}{\delta\sigma(\vv{y})}\nabla\vv{v}\right):\Bigg\{\Lambda_{vv} \left(\nabla\left(\frac{\delta g}{\delta \vv{m}(\vv{y})}\right)-\frac{1}{T}\frac{\delta g}{\delta\sigma(\vv{y})}\nabla\vv{v}\right)
	\nonumber\\
&&\hspace{3.2cm}+\Lambda_{vj} \left(\nabla\left(\frac{1}{\rho}\nabla\times\frac{\delta g}{\delta \vv{B^*}(\vv{y})}\right)-\frac{1}{T}\frac{\delta g}{\delta\sigma(\vv{y})}\nabla\left(\frac{\vv{j}}{\rho}\right)\right)\Bigg\}
	\nonumber\\
	&& +\left(\nabla\left(\frac{1}{\rho}\nabla\times\frac{\delta f}{\delta \vv{B^*}(\vv{y})}\right)-\frac{1}{T}\frac{\delta f}{\delta\sigma(\vv{y})}\nabla\left(\frac{\vv{j}}{\rho}\right)\right)
	\nonumber\\
	&&\hspace{2.5cm}:\Bigg\{ \Lambda_{jv} \left(\nabla\left(\frac{\delta g}{\delta \vv{m}(\vv{y})}\right)-\frac{1}{T}\frac{\delta g}{\delta\sigma(\vv{y})}\nabla\vv{v}\right)
	\label{dispbkt}\\
	&&\hspace{3cm}+\Lambda_{jj} \left(\nabla\left(\frac{1}{\rho}\nabla\times\frac{\delta g}{\delta \vv{B^*}(\vv{y})}\right)-\frac{1}{T}\frac{\delta g}{\delta\sigma(\vv{y})}\nabla\left(\frac{\vv{j}}{\rho}\right)\right)\Bigg\}\Bigg]\,.
	\nonumber
	\end{eqnarray}
When this  bracket of (\ref{dispbkt}) is  subtracted from the  Poisson bracket of (\ref{hamdi}) one obtains the  complete metriplectic geometrical formulation of dissipative XMHD. 
	
	In closing this section let us discuss the forms of the several dissipative tensors. Throughout  this work, we have made no hypotheses on the forms of the various tensors nor on their dependencies on any variables or on  phenomenological coefficients. Our only requirement was that  there be no cross-effects between different tensor-types of responses, a property that  comes from space-parity symmetry \citep{Groot84book}. Yet, physical symmetries will impose other constraints \citep{Groot84book, Landau60book}. Time-reversal symmetry will give the Onsager relations that we have already evoked but not used. Moreover,  Galilean symmetry will constrain the form and dependency of the tensors. Only the magnetic field  can provide directional dependence in the tensors. Anisotropy in Hamiltonian magnetofluids can be introduced by adding a $|\vv{B}|$ dependence to the internal energy $u$ \citep{pjm82,pjmK14}.  Pairing this with anisotropic dissipation would be an  interesting  avenue to explore in the future.   Without anisotropy, the 2-tensors would reduce to scalars while the 4-tensors would decompose into a symmetrization operator, an antisymmetricization operator,  and a trace operator. 
	
	Another constraint is the nonnegativity, which assures  the second law of thermodynamics. If  we  decompose  the various tensors into several scalars, the nonnegativity constraint leads to  nonnegative scalars for direct effects (as distinct from cross effects) and bounds on  the norm of the cross-effect scalars by the geometric means of the two direct-effect scalars of the same kind. 
	
	We have seen that the construction of the brackets does not require physical symmetries, which provides  interesting  insight; \textit{viz.},   the bracket formalism is more general than the  physics at hand. Symmetries only restrict the form of the bracket;  there is  still freedom to select any dependance of the scalars on the phase space and anisotropy due to magnetic directional dependance.

\section{Dissipation in the Lagrangian Picture -- an Example of Metriplectic Reduction}
\label{sec:lag}

\subsection{Lagrangian Picture}

So far,  the formalisms of this paper, both Hamiltonian and dissipative, have been in terms of the Eulerian (spatial) picture of fluid mechanics.  Thus, a natural question  to ask is what would our results look  like in the Lagrangian picture,  where one tracks fluid elements.   For the  Hamiltonian part,  the relationship between the Eulerian and Lagrangian pictures is  well understood for neutral  fluid mechanics \citep[see \textit{e.g.,}][for review]{Morrison98}, magnetohydrodynamics \citep{Morrison09conf}, and XMHD \citep{Charidakos14}.   In the Lagrangian picture one has a canonical Poisson bracket, as expected for a particle-like theory, that reduces to a noncanonical Poisson bracket like the one of equation  (\ref{hamdi})  in the Eulerian picture.  However,  the form of  dissipation in the Lagrangian picture that reduces to the dissipative  bracket  is not evident. Indeed, the lions share of out-of-equilibrium thermodynamics is  studied within  the Eulerian picture.

Let us briefly recall the Lagrangian picture.  In this picture one follows a continuum of  particles, labeled by $\vv{a}$ and then obtains   a flow $\varphi(\vv{a},t)$ that gives the position of the particle, a  fluid element,  labeled by $\vv{a}\in\Omega$ at time $t\in\R$. The configuration space is then the space of the diffeomorphisms of the space $\Omega$. Its cotangent space then defines the momentum $\pi$ and the cotangent bundle will be the  phase space.  In the Hamiltonian setting of the Lagrangian picture, one attaches attributes to a fluid element (see  \textit{e.g.}\ \cite{Morrison09conf}), \textit{viz.}\  mass density $\rho_0(\vv{a})$, entropy density $\sigma_0(\vv{a})$,  and for magnetohydrodynamics, the magnetic field $\vv{B}_0(\vv{a})$.  Then, from $\rho_0$ and $\sigma_0$ we may infer a temperature $T_0(\vv{a})$.  Using the Lagrange to Euler map, the flow is used to obtain the Eulerian velocity field $\vv{v}$ and the attributes are transformed into their well-known Eulerian counterparts that satisfy the  usual equations for the  ideal  fluid and/or magnetohydrodynamics. 

When dissipation is included, we no longer expect attributes to remain independent of time.  For example, the initial entropy 
$\sigma_0(\vv{a},t)$ obtains time dependence, which is consistent with the Eulerian version of this quantity no longer being  conserved in the Eulerian picture. Our goal is to find the Lagrangian equations that determine this time dependence, consistence with our Eulerian metriplectic dynamics. 

 For XMHD the situation is more complicated.  Given that our derivation of section \ref{sec:DXMHD}
starts from  two-fluid theory, we expect there to be two displacement variables. While $\varphi(\vv{a},t)$ will give a center-of-mass displacement, just as for magnetohydrodynamics, we now have $\varphi_d(\vv{a},t)$ that  will evaluate the difference of positions between of ions and electrons of a same label. More precisely, we define $\varphi_d$ as the additional advection of the magnetic field, which will become clearer when we look at  the equations of motion. Conjugate to $\varphi_d$, we have a  momentum variable $\pi_d(\vv{a},t)$.

The Lagrange to Euler map will be  given by the following expressions:  
\bqy
\rho(\vv{x},t)&=&\int_\Omega \ddo{a}  \, \rho_0(\vv{a})  \delta_\Omega\big(\vv{x}-\varphi(\vv{a},t)\big) \,,
\nonumber\\
\sigma(\vv{x},t)&=&\int_\Omega \ddo{a}  \,   \sigma_0(\vv{a},t) \delta_\Omega\big(\vv{x}-\varphi(\vv{a},t)\big) \,,
\nonumber\\
\vv{m}(\vv{x},t)&=&\int_\Omega  \ddo{a}   \,   \pi(\vv{a},t) \delta_\Omega\big(\vv{x}-\varphi(\vv{a},t)\big) \,,
\nonumber
\eqy
where recall $\delta_\Omega$ is the Dirac distribution. Observe, contrary to the usual reduction expressions
 \citep[{\it e.g.},][]{Morrison98,Morrison09conf},  here  the attribute $\sigma_0$  has  explicit time dependence, but since we are not allowing particle production this is not the case for $\rho_0$.   The magnetic field is trickier,  but if the displacement variation $\varphi_d$ is well defined, the magnetic field is advected as a 2-form by $\varphi+\varphi_d$ \citep{DAvignon16}, and we have 
\bq
\vv{B}(\vv{x},t)=\int_\Omega  \ddo a\, (\ddd\varphi+\ddd\varphi_d) \vv{B}_0(\vv{a},t)\, \delta_\Omega\big(\vv{x}-\varphi(\vv{a},t)-\varphi_d(\vv{a},t)\big)\,, 
\nonumber\\
\eq
and again observe $\vv{B}_0$ has explicit time dependence.

From the form of  the Lagrange to Euler map above, we see that given the set of variables $(\varphi,\varphi_d,\pi, \pi_d)$ and known attributes $(\rho_0,\sigma_0,\vv{B}_0)$,  the Eulerian variables are uniquely determined.  However, because of relabeling symmetry and the split between orbit behavior and attribute dynamics, the inverse is not true; \textit{i.e.},  given the Eulerian variables, the Lagrangian $(\varphi,\varphi_d,\pi, \pi_d)$ and attributes $(\rho_0,\sigma_0,\vv{B}_0)$ are not uniquely determined.  Consequently,  like the usual case for Hamiltonian reduction, the  Lagrange to Euler map  is a reduction. Our goal is to find  expressions in the Lagrangian picture that reduce to the known Eulerian equations of the metriplectic dynamical systems that we have described in this paper.  This is an example of metriplectic reduction, an idea that  was introduced in \cite{Materassi18}.

Here we will  choose a particular section that accomplishes metriplectic reduction, even though it  is implicit and  has  a degree of arbitrariness.   In particular, as mentioned above, we will choose a most natural one, where the dissipation changes the attributes, the  fluid element labeled properties,  and not the dynamical displacements. For example,  irreversible processes will make $\sigma_0$ depend on time and increase without altering the form of advection.

%On the other hand, the Eulerian framework fixes space and looks of the property of a fixed point, denoted by $\vv{x}$, in time, whoever the particle in this point is. The system is then described by the mass density $\rho$ which may now depend on time since the space transformation may change the volume, the momentum density $\vv{m}$, the magnetic field $\vv{B}$ and the entropy density $\sigma$. 

\subsection{Relations for  Change of Variables}

In order to find our brackets in the Lagrangian picture, we must use the functional chain rule.  This is done by comparing Lagrangian and Eulerian variations.   However, unlike the usual case of Hamiltonian reduction we include attribute variation. 
As for Hamiltonian reduction,  we have the measure  $\ddo a$ in the Lagrangian picture and  $\ddo x$ in the Eulerian picture. These two volume forms differ by a factor of  the determinant of the flow $\varphi$.   Generalizing  the method of \cite{DAvignon16}, a direct calculation now gives  links between functional derivatives of a functional $f$ of Eulerian variables and its counterpart $\hat{f}$ of Lagrangian variables,   
\bqy
\ddf{\hat{f}}{\pi}(\vv{a},t)&=&\ddf{f}{\vv{m}}(\varphi(\vv{a},t),t)\,,
\nonumber\\
\ddf{\hat{f}}{\sigma_0}(\vv{a},t)&=&\ddf{f}{\sigma}(\varphi(\vv{a},t),t)\,,
\nonumber\\
\ddf{\hat{f}}{\vv{B_0}}(\vv{a},t)&=&(\ddd\varphi+\ddd\varphi_d)^T\ddf{f}{\vv{B}}((\varphi+\varphi_d)(\vv{a},t),t)\,,
\nonumber
\eqy
 where $A^T$ is the transposed operator of $A$.  For XMHD, we need the functional derivative with respect to $\vv{B}^*$ and not $\vv{B}$. A change of variable gives
\[
\ddf{f}{\vv{B}}\longrightarrow \ddf{f}{\vv{B^*}} 
+\mu\chi^2\, \nabla\times\left(\frac{1}{\rho}\nabla\times\ddf{f}{\vv{B^*}}\right)\,.
\]
Using the fact that $\mu$ is of order one, we can invert this relation as a perturbative development in $\mu$. Finally, at order one, we get
\begin{eqnarray*}
	\ddf{f}{\vv{B}^*}(\varphi(\vv{a},t)+\varphi_d(\vv{a},t),t)&\approx&(\ddd\varphi+\ddd\varphi_d)^{-1,T}\ddf{\hat{f}}{\vv{B_0}}(\vv{a},t)\\
	&&\hspace{-.75cm} -\mu\chi^2\nabla\times\left(\frac{1}{\rho}\nabla\times(\ddd\varphi+\ddd\varphi_d)^{-1,T}\ddf{\hat{f}}{\vv{B_0}}(\vv{a},t)\right)\,,
\end{eqnarray*}
 where the gradients are with respect to $\vv{x}=\varphi(\vv{a},t)$. 

To be able to perform the change of variables in the Eulerian bracket, we have to express some variables in terms of Lagrangian ones. One can see that $T$ will become $T_0$ and $\vv{v}$ will become ${\pi}/{\rho_0}$. The Eulerian gradient will transform to the Lagrangian gradient by $\nabla_x f(\varphi(\vv{a}))=(\ddd\varphi)^{-1}\nabla_a \hat{f}(\vv{a})$. Finally, the electric current will become 
\begin{eqnarray*}
		\vv{j}((\varphi+\varphi_d)(\vv{a},t),t)&=&\nabla_x\times\vv{B}((\varphi+\varphi_d)(\vv{a},t),t)\\
		&=&(\ddd\varphi+\ddd\varphi_d)^{-1}\nabla_a\times(\ddd\varphi+\ddd\varphi_d)(\vv{B}_0(\vv{a},t))\,.
			\end{eqnarray*}
From these relations, we have all that is needed to {\it unreduce}, \textit{i.e.},   express the brackets in terms of  the Lagrangian variables.  
	
Upon effecting this procedure, the Hamiltonian part  becomes the canonical bracket,
\[
\{f,g\}=\int_\Omega \ddo {a}\left(\ddf{f}{\varphi}\cdot\ddf{g}{\pi}-\ddf{g}{\varphi}\cdot\ddf{f}{\pi}+\ddf{f}{\varphi_d}\cdot\ddf{g}{\pi_d}-\ddf{g}{\varphi_d}\cdot\ddf{f}{\pi_d}\right) \,.
	\] 
	This result is not surprising given the development of \citet{DAvignon16}, where the  Eulerian bracket is derived from this canonical bracket for extended magnetohydrodynamics.   Let us now turn to the dissipative part, which we will first work out explicitly for hydrodynamics.

\subsection{Lagrangian Dissipation for Hydrodynamics}

To make things simple, in this subsection we will first deal with  hydrodynamics, \textit{i.e.},  we only consider the usual viscosity and heat conductivity, dropping the magnetic part, which is tedious and presents a subtlety that we will address later. Lagrangian metriplectic dynamics was previously explored in \citet{Materassi15};  however, our  study here  adds    the tools needed to address the magnetic part. For hydrodynamics, the change of variables is direct and gives 
\begin{eqnarray}
 (f,g)&=&  \int_\Omega  \ddo {a}\, \frac{T_0}{\T}\Bigg[\nabla\left(\frac{1}{T_0}\frac{\delta f}{\delta\sigma_0(\vv{a})}\right)\cdot T_0^2\, \eta_{T_0T_0} \nabla\left(\frac{1}{T_0}\frac{\delta g}{\delta\sigma_0(\vv{a})}\right)
 \nonumber\\
	&& +\left(\nabla\left(\frac{\delta f}{\delta \pi(\vv{a})}\right)-\frac{1}{T_0}\frac{\delta f}{\delta\sigma_0(\vv{a})}\nabla\left(\frac{\pi}{\rho_0}\right)\right)
	\nonumber\\
	&&\hspace{1cm} 
	:\Lambda_{\pi\pi} \left(\nabla\left(\frac{\delta g}{\delta \pi(\vv{a})}\right)-\frac{1}{T_0}\frac{\delta g}{\delta\sigma_0(\vv{a})}\nabla\left(\frac{\pi}{\rho_0}\right)\right)\Bigg]\,,
	\label{Dlag}
	\end{eqnarray}
where we have supressed the explicit time dependence of  $\pi$ and $\sigma_0$.  Here $\eta_{T_0T_0}= \frac{1}{|\ddd\varphi^{-1}|}(\ddd\varphi^{-1})^T\kappa_{TT}\ddd\varphi^{-1}$ and $\Lambda_{\pi\pi}= \frac{1}{|\ddd\varphi^{-1}|}(\ddd\varphi^{-1})^T\Lambda_{vv}\ddd\varphi^{-1}$, with $|\ddd\varphi|$ the determinant of the endomorphism $\ddd\varphi$ at each point. To be clear, the multiplication here is composition, and for the 4-tensor $\Lambda$, the contraction is with the first index of each pair, the one linked with the gradient. Hence, these tensors change as 2-form densities under  the mapping $\varphi$. 

Observe, the form of (\ref{Dlag}) is the same as that for  the Eulerian picture, so that the equations will remain the same. Yet, the phenomenological tensors change. They will depend on time and reflect the variation of the physical  proximity of labels at nearby   points. Let us  highlight that even with constant scalar phenomenological tensors in the Eulerian picture (a common assumption),  in the Lagrangian picture they will become general time-dependent tensors, unless the displacement $\varphi$ only generates orthogonal transformations, which is coherent physically. Here again, one can see the strength of considering geometrical tools like tensors rather that assuming a particular form like a scalar, which does not exploit the full geometrical structure. 

Adding the purely magnetic terms, as opposed to cross terms,  is also straightforward for magnetohydrodynamics, with or without the Hall term, and for the full XMHD models. Yet, the expressions  are complicated, consequently, we will not write them here.  However, the cross terms, magnetic with nonmagnetic,  bring new ideas that will be explored in subsection \ref{nonlocal}.

\subsection{Nonlocality in the Lagrangian Picture}
\label{nonlocal}
 
Consider now  the magnetic cross effects.  While the other variables change with $\varphi$ or $\varphi+\varphi_d$ directly, allowing an easy change of variables, the cross-effect between a magnetic variable and a nonmagnetic variable change into one  of the responses (\textit{cf.},  table \ref{table}) with $\varphi$ while the response changes with $\varphi+\varphi_d$.  Because   there is a product between them,   changing variables brings complications. Indeed, we will see that this complication  breaks locality in the Lagrangian picture. But first, let us take a look at our general bracket once more. 

Recall, for  a general nonequilibrium system we saw in  section \ref{sec:bracket} that we have a dissipative bracket of equation (\ref{genBkt}). This bracket can be rewritten as follows: 
\[
(f,g)=\frac{1}{\T}\int_\Omega\ddo{x}\int_\Omega\ddo{y}\, \ddf{f}{\zeta_\alpha(\vv{x})}\mathcal{L}_{\alpha\beta}(\vv{x},\vv{y})\ddf{g}{\zeta_\beta(\vv{y})}\,,
\]
where $\mathcal{L}_{\alpha\beta}(\vv{x},\vv{y}):= \nabla_x\nabla_y\left(L_{\alpha\beta}\delta_\Omega(\vv{x}-\vv{y})\right)$. Note, $L$ could also depend explicitly  on  space and on any variable of the phase space but,  to be concise, we do not exhibit these dependencies. A generalization of the bracket could then be to allow $\mathcal{L}$ to be a general tensor of distributions. So,  why do we have such a special form? First, having only  gradient  factors $\nabla_x\nabla_y$ in this distribution is roughly the linear response assumption of out-of-equilibrium thermodynamics. Second, assuming that we have such a Dirac distribution means that the interactions between the variations of $f$ and $g$ exist only at  the same point. Saying it another way, the value $(f,g)(\vv{x})$ depends only on the values of $f$ and $g$ in an arbitrarily  small neighborhood of $\vv{x}$. This assumption amounts to the assumption of  the locality of the interactions. 

The complication with cross-terms between magnetic and nonmagnetic terms in XMHD arises because   different factors in a term  do not  transform with the same displacement. One  is transformed  by $\varphi+\varphi_d$ while the other is transformed  by $\varphi$. Then, what change of variables do we do? Actually, both. The idea is to express the bracket in its more general form, with two integrals, and to change both variables with their  associated displacements. For simplicity, we will only show how this works for  the thermoelectric effect, which has  the following bracket cross term: 
\[
(f,g)_{\mathfrak{T}}:=\int_\Omega\!  \ddo {y}\frac{T}{\T}\nabla\left(\frac{1}{T}\frac{\delta f}{\delta\sigma(\vv{y})}\right)\!\cdot T\eta_{Tj} \left(\!\nabla\times\!\left(\frac{\delta g}{\delta \vv{B^*}(\vv{y})}\right)-\frac{1}{T}\frac{\delta g}{\delta\sigma(\vv{y})}\vv{j}\right)\,.
\]
The other terms can be treated similarly. 

With the two integrals and a change of the left coordinate using $\varphi$ and the right coordinate using  $\varphi+\varphi_d$,  the bracket becomes the following  in the Lagrangian picture:
\begin{eqnarray*}
(f,g)_{\mathfrak{T}}&=&\int_\Omega  \ddo {a}\int_\Omega  \ddo {b}\, \frac{T_0(\vv{a})T_0(\vv{b})}{\T}\nabla_a\left(\frac{1}{T_0(\vv{a})}\frac{\delta f}{\delta\sigma_0(\vv{a})}\right)\cdot \eta_{T_0 j_0} \\
&&\Bigg[\nabla_b\times\Bigg((\ddd\varphi+\ddd\varphi_d)^{-1,T}\ddf{f}{\vv{B_0}}(\vv{b})
\\
	&&%\hspace{-.5cm}
	 -\mu\chi^2\nabla_b\times\left(\frac{1}{\rho}(\ddd\varphi+\ddd\varphi_d)^{-1}\nabla_b\times(\ddd\varphi+\ddd\varphi_d)^{-1,T}\ddf{f}{\vv{B_0}}(\vv{b})\right)\Bigg)\\
	&&-\frac{1}{T_0(\vv{b})}\frac{\delta g}{\delta\sigma_0(\vv{b})}\nabla_b\times(\ddd\varphi+\ddd\varphi_d)(\vv{B}_0(\vv{b}))\Bigg]\,,
	\end{eqnarray*}
where
\[
\eta_{T_0 j_0}=(\ddd\varphi^{-1})^T\kappa_{Tj}\ddd(\varphi+\varphi_d)^{-1}\frac{\delta_\Omega(\varphi(\vv{a})-(\varphi+\varphi_d)(\vv{b}))}{|\ddd\varphi^{-1}||\ddd(\varphi+\ddd\varphi_d)^{-1}|}\,.
\]

The important point to realize here is that if $\varphi_d$ does vanish, then the bracket reduces to a bracket of the same form as that  for hydrodynamics. But if $\varphi_d$ does not vanish, then the locality assumption breaks. How should this be interpreted? Well, $\varphi_d$ is roughly the inertia of the electrons. Thus, this nonlocality is saying that ions and electrons located at the  same space point will interact (locality in the Eulerian picture) but that these two kinds of particles do not come from the same label, for they do not have the same dynamics (nonlocality in the Lagrangian picture). Thus, for magnetohydrodynamics and even Hall magnetohydrodynamics, where electron inertia is neglected, locality is saved in the Lagrangian picture; in the equations, $\varphi_d$ is identically zero. On the other hand, in XMHD, locality is broken in the Lagrangian picture. 

It is interesting to see how a   more general bracket, which might have appeared useless,  appears naturally in a physical system. Studying more precisely the consequences of such a nonlocality would be a  useful avenue for  future work. This study also sheds light on the physical consequences of electron inertia.

\section{Conclusion}\label{sec:conclusion}
%\subsection{Outlook}

In this paper, we have derived a conservative yet dissipative form of XMHD from two-fluid theory.  We have seen that natural dissipation and cross-effects appear, including  a new current viscosity. We have seen that this current viscosity is small, explaining why it is mostly neglected. Yet, we have explained it physically and described consequences  of its associated cross effects.

We have also constructed a general metriplectic framework for any fluid-like nonequilibrium thermodynamic system and presented a systematic way to derive the dissipative brackets. The main new idea was to use conserved  thermodynamic variables, which are natural in this context but differ from the usual Hamiltonian variables, which explains why they are not usually used. As an example, we re-discovered and generalized the hydrodynamic bracket of \citet{Morrison84b} using  this new framework. 

With the hydrodynamic experience, we derived for the first time the  metriplectic bracket for full dissipative XMHD. We also explained the geometric generality of our result, freeing us from any dependence on  the phase space variables or the direction of the magnetic field, thereby obtaining  more general  equations than typically  used for this model. 

Finally, we used these geometrical tools to study this model in the Lagrangian picture. In this picture, we still have a natural bracket, but two generalizations appear naturally.  First, the geometry of the phenomenological tensor becomes general and time-dependent. This allows for the description when scalar phenomenological tensors are no longer a good approximation. Second, the locality assumption can break, and then we must  consider a more general form of dissipative bracket. This occurs for  XMHD, because the  ions and electrons have separate dynamics.

\section*{Acknowledgements}

BC would like to acknowledge the hospitality of the Institute for Fusion Studies and the Department of Physics of The University of Texas at Austin.  PJM  was supported by  the U.S. Department of Energy Contract DE-FG05-80ET-53088 and via a  Forschungspreis from the Humboldt Foundation.    He warmly acknowledge the hospitality of the Numerical Plasma Physics Division of Max Planck IPP, Garching, where a portion of this research was done.

%Bibliography
\bibliography{bibfile}
\bibliographystyle{jpp}

\end{document}